\documentclass[11pt]{amsart}
\usepackage{etex}
\newtheorem{Theorem}{Theorem}[section]
\newtheorem{Proposition}[Theorem]{Proposition}
\newtheorem{Lemma}[Theorem]{Lemma}

\newtheorem{Example}[Theorem]{Example}
\newtheorem{Remark}[Theorem]{Remark}

\newtheorem{Assumption 2}[Theorem]{Assumption 2}

\numberwithin{equation}{section}

\usepackage{amsmath}
\usepackage{amssymb}
\usepackage{lmodern}
\usepackage{stmaryrd}

\let\savedegree\degree
\let\degree\relax
\usepackage{mathabx}
\let\degree\savedegree

\usepackage{gensymb}

\def\k#1{\kern#1em}
\def\Ib#1{{I\kern-.25em#1}}
\def\Ibb#1{{I\kern-.23em#1}}

\def\CC{{\mathbb C}}

\def\NN{{\mathbb N}}

\def\RR{{\mathbb{R}}}

\def\vci{\vrule  width.02em height1.47ex depth-.0ex}
\def\11{{\rm\k{.2}\vci\k{-.37}1}}

\def\fin{{\begin{flushright}
\it{Q.E.D.}
\end{flushright}}}

\begin{document}

\address{Universit\'e de Bordeaux, Institut de Math\'ematiques, UMR CNRS 5251, F-33405 Talence Cedex}

\email{alain.bachelot@u-bordeaux.fr}

\title{Propagation of Massive Scalar Fields in Pre-Big Bang Cosmologies}

\author{Alain BACHELOT}

\begin{abstract}
We investigate the linear and semilinear massive Klein-Gordon
equations in geometrical frameworks of type ``Conformal Cyclic
Cosmology'' of R. Penrose, or ``Singular Bouncing Scenario'' as well. We give sufficient conditions on the decay of the mass to the fields be able to propagate across the Big-Bang.
\end{abstract}

\maketitle

\pagestyle{myheadings}
\markboth{\centerline{\sc Alain Bachelot}}{\centerline{\sc {{Propagation of Massive Scalar Fields in Pre-Big Bang Cosmologies}}}}

\section{Introduction}
In this work we investigate the propagation of a {\it massive} scalar field in
simple models of the pre-big bang cosmology, in particular for  the Conformal Cyclic Cosmology (CCC). In this scenario
developped by R. Penrose \cite{ccc}, \cite{penrose2018}  (see also \cite{camara},
\cite{lubbe}, \cite{meissner}, \cite{newman}, \cite{tod}), we consider a $n+1$ dimensional $C^2$, time oriented, Lorentzian manifold
$(\mathcal{M},g)$. We assume that this spacetime is
globally hyperbolic. Given  a Cauchy hypersurface $\Sigma$ we split
$\mathcal{M}$ as
\begin{equation}
 \label{}
 \mathcal{M}=\hat{\mathcal{M}}\cup \Sigma\cup\check{\mathcal{M}},\;\; \hat{\mathcal{M}}:=I^-(\Sigma),\;\;\check{\mathcal{M}}:=I^+(\Sigma),
\end{equation}
where $I^-(\Sigma)$ (resp. $I^+(\Sigma)$) is the chronological past
(resp. chronological future) of $\Sigma$.
The past (resp. future) set $\hat{\mathcal{M}}$
(resp. $\check{\mathcal{M}}$) is an open subset
called {\it Previous Aeon} (resp. {\it Present Aeon}). The metric
$g$ is just a mathematical tool, called {\it bridging metric}, and the physical metrics $\hat{g}$
and $\check{g}$ on the Aeons
are solutions of the Einstein equations that are conformally
equivalent to $g$,
\begin{equation}
 \label{gg}
 \hat{g}=\hat{\Omega}^2g,\;\;\check{g}=\check{\Omega}^2g,
\end{equation}
where $\bar{\Omega}$ is a non-vanishing $C^2$ function on
$\bar{\mathcal{M}}$ (throughout the paper, the bar stands for the hat
or the check). In the CCC we assume that $\check{\Omega}$ can be
continuously extended by zero to $\Sigma$. Therefore the Present Aeon
is an universe beginning by a Big Bang, and $\Sigma$ is called the {\it
  Bang Surface}. In contrast we assume that $\hat{\Omega}$ tends to
the infinity at $\Sigma$ that is the Future Infinity of
the Previous Aeon. In this paper we consider the simple but
important case, of the static metric $g$, therefore the Aeons are
generalized FLRW spacetimes. Our results can also be applied to the
{\it Singular Bouncing Scenario} (see {\it e.g} \cite{battefeld},
\cite{brandenberger}, \cite{gasperini}) for which $\hat{\Omega}$ and $\check{\Omega}$
tend to zero at $\Sigma$, and even for the opposite case of an
expansion-contraction transition,
where $\Sigma$ is the Future Infinity of $\hat{\mathcal{M}}$ and the
Past Infinity of $\check{\mathcal{M}}$ if $\hat{\Omega}$ and $\check{\Omega}$
tend to infinity at $\Sigma$ (at our knowledge this extreme case had
not been considered yet). We address the fundamental issue : to
find sufficient conditions on the decay of the mass near $\Sigma$ so that a
massive scalar field in $\hat{\mathcal{M}}$ can be extended
in $\check{\mathcal{M}}$ despite the severe singularity on
$\Sigma$ that leads to the blow-up or the vanishing of the fields on
it. The idea is following: we renormalize the fields $\bar{u}$ by
using the Liouville transforms associated to the conformal factors
\begin{equation}
 \label{liou}
 \bar{\varphi}:=\bar{\Omega}^{\frac{n-1}{2}}\bar{u},
\end{equation}
then we investigate the wave equation satisfied by the renormalized
fields $\bar{\varphi}$. We impose some constraints on the mass
that allow to obtain the existence of trace of $\bar{\varphi}$ and a suitable
time-derivative, on the
  bang surface $\Sigma$. We conclude that the asymptotic behaviour
  of $\hat{u}$ at $\Sigma$ in the Previous Aeon, leads to a Cauchy data  that determines
  $\check{u}$ in the Present Aeon.

A free massive scalar field obeys the Klein-Gordon
equation satisfied in the interior $\ring{\bar{\mathcal{M}}}$ of each
Aeon $\bar{\mathcal{M}}$,
\begin{equation}
\label{kg}
 \left[
   {\Box}_{\bar{g}}+\bar{\xi} R_{\bar{g}}+\bar{m}^2\right]\bar{u}=\bar{f}\;\;in\;\;\ring{\bar{\mathcal{M}}}.
\end{equation}
Here ${\Box}_{h}$ and $R_h$ are respectively the D'Alembert operator
and the Ricci scalar associated to a metric $h$,
\begin{equation}
 \label{}
  {\Box}_{h}u:=\frac{1}{\sqrt{|\det(h)|}} \partial_{\mu}\left(h^{\mu \nu}
\sqrt{|\det(h)|}\partial_{\nu}u\right),
\end{equation}
$\bar{\xi}$ is a constant describing the coupling with the geometry, the mass $\bar{m}$ is a non negative 
function defined on $\bar{\mathcal{M}}$ and the source term $\bar{f}$
is given. The fundamental issue
concerns the propagation of the field from the Previous Aeon to the
Present Aeon across the Bang Surface. The true problem is fully non
linear and we should deal with the coupled Einstein-Scalar Field
system.
A very important result in this domain is due to H. Friedrich \cite{fried}
who investigated the 1+3 dimensional  case with a cosmological
constant $\Lambda>0$ and
\begin{equation}
 \label{fried}
 \hat{\xi}=0,\;\;\hat{m}^2=\frac{2}{3}\Lambda.
\end{equation}
Our aim is more modest : we study the linear or semilinear Klein-Gordon equation in
a fixed geometrical framework of the CCC or the Bouncing Scenario (semi-classical approximation for the
weak fields). In this context, there is a trivial situation, called the
{\it conformal invariant massless case}, defined by
\begin{equation}
 \label{xic}
 \bar{\xi}=\frac{n-1}{4n},\;\;
 \bar{m}=0.
\end{equation}
Using the Liouville transform (\ref{liou})
we can easily check that $\bar{u}$ is a solution of (\ref{kg}) iff $\bar{\varphi}$ is solution of
the Klein-Gordon equation with the variable effective mass
$\mid \bar{m}\bar{\Omega}\mid$, associated to the bridging metric $g$,
\begin{equation}
 \label{kgt}
 \left[  {\Box}_{g}+\bar{\xi} R_{g}+n\bar{\Omega}^{-1}\left(\bar{\xi}-\frac{n-1}{4n}\right)\left(2\bar{\Omega}_{;\mu\nu}+(n-3)\bar{\Omega}^{-1}\bar{\Omega}_{;\mu}\bar{\Omega}_{;\nu}\right)g^{\mu\nu}+\bar{m}^2\bar{\Omega}^2\right]\bar{\varphi}=\bar{\Omega}^{\frac{n+3}{2}}\bar{f}.
\end{equation}
Hence for the  conformal invariant massless case (\ref{xic}) with $\bar{f}=0$, the propagation of the field is just
described by the wave equation
\begin{equation}
 \label{kdo}
 \left[  {\Box}_{g}+\frac{n-1}{4n}R_{g}\right]\bar{\varphi}=0,
\end{equation}
and since $(\mathcal{M},g)$ is globally hyperbolic, the global Cauchy
problem is well posed for this equation by the Leray theorem : given a
Cauchy hypersurface, any initial data on it determines a unique
solution defined on the whole manifold. In this sense, the
conformal invariant massless case without source term is trivial: the field freely propagates from
the Previous Aeon to the Present Aeon.\\

The situation drastically
changes if $\bar{\xi}\neq (n-1)/4n$ or $\bar{m}\neq 0$ since
$\bar{\Omega}^{-1}$ (resp $\bar{\Omega}$) blows up on $\Sigma$ for
a Big Bang (resp. an expanding universe\footnote{However when
  $\hat{g}$ is a solution of the Vacuum Einstein Equations
  $R_{\mu\mu}-\frac{1}{2}Rg_{\mu\mu}+\Lambda g_{\mu\mu}=0$, we have
  $R_{\hat{g}}=2\frac{n+1}{n-1}\Lambda$, and the case
  $(\xi=0,m^2=\frac{n+1}{2n}\Lambda)$ is equivalent to the conformal
  invariant massless case. In particular Friedrich in \cite{fried}
  studied a
non-linear version of  (\ref{kdo}) as well as the coupling to the background, which in the present article is prescribed.}) and the Klein-Gordon equation (\ref{kgt}) is highly
singular. As a consequence $\bar{\varphi}$ can diverge at $\Sigma$ and
the possibility of a propagation from the Previous Aeon to the Present
Aeon seems to be doubtful in general. For instance if $\hat{\xi}=0$,
$\hat{m}$ is a strictly positive constant and
$\hat{g}=\tau^{-2}\left(d\tau^2-g_{ij}d\mathrm{x}^{i}d\mathrm{x}^j\right)$ is De-Sitter
like, the results established by A. Vasy in \cite{vasy} show that
$\hat{\varphi}(\tau,.)\sim
\tau^{\frac{1}{2}-\sqrt{\frac{n^2}{4}+\hat{m}^2}}\psi(\tau,.)$ near
$\Sigma=\{\tau=0\}$ with
$\psi\in C^0(\hat{\mathcal{M}}\cup\Sigma)$ (see also \cite{dodeca} for
$\hat{\xi}=\hat{m}=0$). In this paper we follow
the ideas of Penrose and we assume that $\bar{\xi}=(n-1)/4n$ and the
mass of the field is a function that enjoys sufficient integrability
with respect to time near $\Sigma$. Therefore we investigate the Klein-Gordon
equations
\begin{equation}
 \label{KG}
 \left[  {\Box}_{g}+\frac{n-1}{4n}R_{g}+\bar{m}^2\bar{\Omega}^2\right]\bar{\varphi}=\bar{\Omega}^{\frac{n+3}{2}}\bar{f}\;\;\;in\;\;\ring{\bar{\mathcal{M}}},
\end{equation}
and we look for sufficient conditions on $\bar{m}$ to that
$\hat{\varphi}$  determines $\check{\varphi}$ despite the
singularity on $\Sigma$.
An obvious way could consist in introducing
\begin{equation}
 \label{deftilde}
 \tilde{m}=\hat{m},\;\;\tilde{\Omega}=\hat{\Omega},\;\;\tilde{f}=\hat{f}\;\;on
\;\;\hat{\mathcal{M}},\;\;\tilde{m}=\check{m},\;\;
\tilde{\Omega}=\check{\Omega},\;\;\tilde{f}=\check{f}\;\;on\;\;\check{\mathcal{M}},
\end{equation}
and solving on
the whole manifold $\mathcal{M}$ the equation
\begin{equation}
 \label{KGbof}
 \left[  {\Box}_{g}+\frac{n-1}{4n}R_{g}+\tilde{m}^2\tilde{\Omega}^2\right]\tilde{\varphi}=\tilde{\Omega}^{\frac{n+3}{2}}\tilde{f}\;\;\;in\;\;\ring{\mathcal{M}}.
\end{equation}
Then $\check{\varphi}$ could be defined by $\tilde{\varphi}$ on
$\check{\mathcal{M}}$ if $\tilde{\varphi}=\hat{\varphi}$ on
  $\hat{\mathcal{M}}$. This elementary approach could work if the
  equation (\ref{KGbof}) makes sense. In particular
  $\tilde{m}^2\tilde{\Omega}^2 \tilde{\varphi}$ has to be well defined as a
  distribution in $\mathcal{M}$. Unfortunately
  $\hat{m}^2\hat{\Omega}^2$ is not in $L^1_{loc}(\hat{\mathcal{M}}\cup\Sigma)$ in
  general, unless we add unreasonably strong conditions on the decay of the
  mass. In fact (\ref{KGbof}) holds for the Singular Bouncing
  Scenario for which $\tilde{m}^2\tilde{\Omega}^2\in L^1_{loc}(\mathcal{M})$ (see Theorem \ref{teofastoch} and Theorem
  \ref{teofastochnl}), but this very simple method fails for the CCC
  for which $\tilde{m}^2\tilde{\Omega}^2\notin L^1_{loc}(\mathcal{M})$
  and equation (\ref{KGbof}) does not make sense. To overcome this
  difficulty and to obtain the main results of this paper (Theorem
  \ref{teodur} and Theorem \ref{teodurnl}), we adapt a method used in
  \cite{RIP} and \cite{delsanto} based on the solving of the Riccati
  equation
  $$
  \bar{A}'-\bar{A}^2= \bar{m}^2\bar{\Omega}^2
  $$
  that allows to transform equation (\ref{KG}) into an new equation
  in which all the coefficients are integrable functions (see (\ref{ekcefifi})
  and \ref{ekcefifika})). From a mathematical point of view, this work
  deals with a linear scalar wave equation with a time-dependent mass, or in
  a
  time-dependent background, a
  topic for which a lot of papers have been devoted, see {\it e.g.}
  \cite{ebert2017} and the references therein, and in the framework of
  General Relativity \cite{RIP}, \cite{dodeca}, \cite{ringstrom2017},
  \cite{ringstrom2019}, \cite{vasy}. We also investigate the
  semi-linear Klein-Gordon equation for $n=3$. This important model has
  been investigated in the case of the smooth Lorentzian manifolds and
  fixed mass, in
  particular in \cite{cagnac}, \cite{ebert2018}, \cite{galstian2015},
  \cite{galstian2017}, \cite{joudioux}, \cite{nakamura},
  \cite{nicolas1995}, \cite{nicolas2002}; here we consider the
  semi-linear equation (\ref{KG}) with $\bar{f}=-\kappa\mid
 \bar{u}\mid^2\bar{u}$, and the strongly singular effective mass
 $\bar{m}\bar{\Omega}$, and we prove that, like for the linear case,
 the fields propagate across the Bang Surface.\\

We end this introduction by some remarks from the perspective of
Physics.
The fundamental issue is the stability of the
 Einstein-scalar field system for a cosmological scenario with a pre-Big
 Bang era: could a disturbance in the Past Aeon dramatically affect  the Present
 Aeon?
 Unfortunately we are an extremely long way from achieving
 this ultimate goal. First of all, no clear physical theory describing a physically
 meaningful transition from the Past Aeon to the Present Aeon,
 exists yet. Among several interesting tentatives, we can cite,
 besides the CCC, the
 String Cosmology of Gasperini and Veneziano \cite{gasperini} (for a
 large overview, see \cite{brandenberger}), and the extensions of the
 FLRW models to negative times for which the Bang surface is light-like \cite{klein}. Secondly, investigating the
 Einstein equations, we face
 considerable mathematical difficulties. There are important results
 of stability for the standard cosmology with a single Aeon
 beginning with a Big-Bang (in this context we have to mention
 \cite{ringstrom2008}, \cite{ringstrom2013}, \cite{rodnianski2018} and
 \cite{rodnianski2018-1}). The CCC is a much more problematic scenario, but the
 deep results of Friedrich \cite{fried}, \cite{fried2017} are very
 promising. Here, we ignore the Einstein equations, and our work falls
 within the framework of the field theory since we investigate the propagation
 of the scalar fields in a fixed geometrical setting. The semilinear
 Klein-Gordon equation with a dispersive non-linearity that we
 consider,  is a kind of master equation for several particles : Higgs
 boson,  {\it Inflaton} of the inflationay epoch, Brans-Dicke
 scalar field, and in the CCC,  {\it Erebon} that is a dark-matter particle
 introduced by Penrose, solution of a Yamabe equation. We let open the
 much more difficult case of the quasilinear equations
 $\Box_{\bar{g}}\bar{u}=Q(x,\bar{u},\nabla \bar{u}, \nabla^2\bar{u})$ that could be a preliminary step
 toward the Einstein equations.
\section{Geometrical and Functional frameworks}

We consider a
$n$-dimensional $C^{\infty}$ Riemannian manifold $(\mathbf
K,\gamma)$, $n\geq 1,$ and $\tau_-,\tau_+$, $-\infty<\tau_-<0<\tau_+<\infty$.
The Lorentzian manifold $\mathcal{M}$ and the bridging metric $g$ are
given by:
\begin{equation}
 \label{mani}
 \mathcal{M}=[\tau_-,\tau_+]_{\tau}\times\mathbf{K}_{\mathrm x},\;\;g_{\mu\nu}dx^{\mu}dx^{\nu}=d\tau^2-\gamma_{ij}d\mathrm{x}^id\mathrm{x}^j,
\end{equation}
and we take
\begin{equation}
 \label{}
 \Sigma:=\{\tau=0\}\times\mathbf{K},\;\;\hat{\mathcal{M}}:=[\tau_-,0)\times\mathbf{K},\;\; \check{\mathcal{M}}:=(0,\tau_+]\times\mathbf{K}.
\end{equation}
Given two non-vanishing functions $\hat{\Omega}\in
C^2(\hat{\mathcal{M}})$, $\check{\Omega}\in
C^2(\check{\mathcal{M}})$ we  endowe $\bar{\mathcal{M}}$ with the
metrics defined by (\ref{gg}). Some of our results are obtained without
 additional assumptions on $\bar{\Omega}$. Nevertheless the
cases of physical interest are (1) the {\it Singular Bouncing Scenario} with
\begin{equation}
 \label{bounce}
 \forall
 \mathrm{x}\in\mathbf{K},
 \;\;\hat{\Omega}(\tau,\mathrm{x})\xrightarrow[\tau\to 0^-]{}
 0,\;\;\check{\Omega}(\tau,\mathrm{x})\xrightarrow[\tau\to 0^+]{}
 0,
\end{equation}
($\tau=0$ is a Big Crunch for $\hat{\mathcal{M}}$ and a Big Bang for $\check{\mathcal{M}}$),
and (2) the CCC of Penrose for which
\begin{equation}
 \label{}
 \forall
 \mathrm{x}\in\mathbf{K},\;\;\int_{\tau_-}^0\mid\hat{\Omega}(\tau,\mathrm{x})\mid d\tau=\infty,
\;\;\check{\Omega}(\tau,\mathrm{x})\xrightarrow[\tau\rightarrow 0^+]{}
 0,
\end{equation}
($\tau=0$ is the Future Infinity for $\hat{\mathcal{M}}$ and a Big
Bang for $\check{\mathcal{M}}$). We could also consider more
extreme cases such as an infinite expansion-contraction with
\begin{equation}
 \label{}
 \forall
 \mathrm{x}\in\mathbf{K},\;\;\int_{\tau_-}^0\mid\hat{\Omega}(\tau,\mathrm{x})\mid
 d\tau=\int_0^{\tau_+}\mid\check{\Omega}(\tau,\mathrm{x})\mid
 d\tau=\infty,
\end{equation}
for which $\tau=0$  is the Future Infinity for $\hat{\mathcal{M}}$ and
the Past Infinity for $\check{\mathcal{M}}$, or a Previous Aeon ending
with a {\it Big Rip} at $\tau=0$,
\begin{equation}
 \label{}
 \forall
 \mathrm{x}\in\mathbf{K},\;\;\int_{\tau_-}^0\mid\hat{\Omega}(\tau,\mathrm{x})\mid
 d\tau<\infty,\;\; \hat{\Omega}(\tau,\mathrm{x})\xrightarrow[\tau\rightarrow 0^-]{}
 \infty.
\end{equation}

We shall be able to apply our results to a very important example of
CCC that is associated to generalized FLRW
universes:
\begin{Example}
 \label{}
Given  $\hat{t}_-\in\RR$,
we introduce
\begin{equation}
 \label{}
 \hat{\mathcal{M}}:=[\hat{t}_-,\infty)_{\hat{t}}\times\mathbf{K}_{\mathrm
   x},
\end{equation}
\begin{equation}
 \label{}
\hat{g}_{\mu\nu}dx^{\mu}dx^{\nu}=d\hat{t}^2-\hat{a}(\hat{t})^2\gamma_{ij}d\mathrm{x}^id
\mathrm{x}^{j}.
\end{equation}
Here the scale factor $\hat{a}$ is a strictly positive function in
$C^2([\hat{t}_-,\infty))$. We assume that the expansion is sufficiently
accelerating to that
\begin{equation}
 \label{}
 \hat{a}^{-1}\in L^1(\hat{t}_-,\infty).
\end{equation}
Now we introduce  the conformal time $\tau$ that is defined as
\begin{equation}
 \label{}
 \tau(\hat{t}):=-\int_{\hat{t}}^{\infty}\frac{1}{\hat{a}(s)}ds
\end{equation}
and in the $(\tau,\mathrm{x})$ coordinates, the Previous Aeon is described
as
\begin{equation}
 \label{}
 \hat{\mathcal{M}}=[\tau_-,0)_{\tau}\times\mathbf{K}_{\mathrm x},\;\;\tau_-:=-\int_{\hat{t}_-}^{\infty}\frac{1}{\hat{a}(s)}ds\in(-\infty,0),
\end{equation}
\begin{equation}
 \label{}
\hat{g}_{\mu\nu}dx^{\mu}dx^{\nu}=\hat{\Omega}^2(\tau)\left[d\tau^2-\gamma_{ij}d\mathrm{x}^id
\mathrm{x}^{j}\right],\;\;\hat{\Omega}(\tau):=\pm\hat{a}(\hat{t}).
\end{equation}
An important particular case is the De-Sitter like metric for which
\begin{equation}
 \label{ds}
 \hat{a}(\hat{t})\sim \hat{C} e^{\hat{H}\hat{t}},\;\;\hat{t}\rightarrow\infty,\;\;\hat{C},\hat{H}>0.
\end{equation}
We easily check that
\begin{equation}
 \label{}
 \hat{\Omega}(\tau)\sim \mp\frac{1}{\hat{H}\tau},\;\;\tau\rightarrow 0^-.
\end{equation}

In a similar way, given $\check{t}_+\in(0,\infty)$
we introduce
\begin{equation}
 \label{}
 \check{\mathcal{M}}:=(0,\check{t}_+]_{\check{t}}\times\mathbf{K}_{\mathrm
   x},
\end{equation}
\begin{equation}
 \label{}
\check{g}_{\mu\nu}dx^{\mu}dx^{\nu}=d\check{t}^2-\check{a}(\check{t})^2\gamma_{ij}d\mathrm{x}^id
\mathrm{x}^{j}.
\end{equation}
Now the scale factor $\check{a}$ is a strictly positive function in
$C^2((0,\check{t}_+])$ that tends to zero as $\check{t}\rightarrow 0$
sufficiently slowly to that
\begin{equation}
 \label{}
 \check{a}^{-1}\in L^1(0,\check{t}_+).
\end{equation}
The conformal time $\tau$ is defined as
\begin{equation}
 \label{}
 \tau(\check{t}):=\int_0^{\check{t}}\frac{1}{\check{a}(s)}ds
\end{equation}
and in the $(\tau,\mathrm{x})$ coordinates, the Present Aeon is described
as
\begin{equation}
 \label{}
 \check{\mathcal{M}}=(0,\tau_+]_{\tau}\times\mathbf{K}_{\mathrm x},\;\;\tau_+:=\int_{0}^{\check{t}_+}\frac{1}{\check{a}(s)}ds\in(0,\infty),
\end{equation}
\begin{equation}
 \label{}
\check{g}_{\mu\nu}dx^{\mu}dx^{\nu}=\check{\Omega}^2(\tau)\left[d\tau^2-\gamma_{ij}d\mathrm{x}^id
\mathrm{x}^{j}\right],\;\;\check{\Omega}(\tau):=\check{a}(\check{t}).
\end{equation}
An interesting case is the $C^0$ Big Bang studied in \cite{RIP} for
which
\begin{equation}
 \label{bb}
 \check{a}(\check{t})\sim \check{C}\check{t}^{\check{\eta}},\;\;\check{t}\rightarrow
 0^+,\;\;0<\check{C},\;\;\check{\eta}\in (0,1).
\end{equation}
Then we have
\begin{equation}
 \label{}
 \check{\Omega}(\tau)\sim
 \check{C}^{\frac{1}{1-\check{\eta}}}(1-\check{\eta})^{\frac{\check{\eta}}{1-\check{\eta}}}\tau^{\frac{\check{\eta}}{1-\check{\eta}}},\;\;\tau\rightarrow 0^+.
\end{equation}
In particular, $\check{\eta}=2/3$, $n=3$, $\mathbf{K}=\mathbb{T}^3$ is the famous FLRW solution of which
the stability is investigated in \cite{rodnianski2018}, \cite{rodnianski2018-1}.
We remark that in the cases (\ref{ds}), (\ref{bb}), with
$\tau_-=-\tau_+$, the map $\tau\mapsto-\tau$ relates
$\hat{\mathcal{M}}$ and $\check{\mathcal{M}}$. Then the Penrose's ``reciprocal proposal'' \cite{ccc}
\begin{equation}
 \label{}
 \hat{\Omega}(-\tau)\check{\Omega}(\tau)=-1,\;\;\tau\in (0,\tau_+),
\end{equation}
can be satisfied when
\begin{equation}
 \label{}
 \hat{\Omega}(\tau)=-\hat{a}(\hat{t}),\;\;\check{\eta}=\frac{1}{2},\;\;\check{C}=\sqrt{2\hat{H}}.
\end{equation}

\end{Example}

As regards the manifold $(\mathbf{K},\gamma)$ we assume it is complete
and the scalar curvature $R_{\gamma}$ is bounded
\begin{equation}
 \label{rb}
 R_{\gamma}\in L^{\infty}(\mathbf{K}).
\end{equation}
Here we denote $L^p(\mathbf K)$ the $L^p$-Lebesgue space on $\mathbf{K}$
endowed with the volume measure associated to the metric
$\gamma$. It is well known the the Laplace-Beltrami operator
$$
\Delta_{\mathbf K}:=\frac{1}{\sqrt{|\det(\gamma)|}} \partial_{i}\left(\gamma^{ij}
\sqrt{|\det\gamma)|}\partial_{j}u\right),
$$ 
is essentially selfadjoint on
$C^{\infty}_0(\mathbf K)$. Hence we can introduce the Sobolev spaces $H^s(\mathbf K)$,
$s\in\RR$, defined as the closure of $C^{\infty}_0(\mathbf K)$ for the norm
\begin{equation}
 \label{normhs}
 \Vert f\Vert_{H^s(\mathbf K)}:=\left\Vert \left(-\Delta_{\mathbf
       K}+1\right)^{\frac{s}{2}}f\right\Vert_{L^2(\mathbf K)}.
\end{equation}

To investigate the semi-linear Klein-Gordon equation, we need the
Sobolev embedding, therefore we have to strengthen the
assumptions on the metric. Thus to study the self-interacting fields,
we shall assume that $(\mathbf{K},\gamma)$ is a $C^{\infty}$ {\it
  bounded geometry manifold}, {\it i.e.} the
following conditions are satisfied: (1) the injectivity radius is
strictly positive, (2) every covariant derivative of the Riemannian
curvature tensor is bounded, or in an equivalent way,
\begin{equation}
 \label{bg}
 \forall\alpha\in\NN^n,\;\;D^{\alpha}\gamma_{ij}\in L^{\infty}(\mathbf{K}),\;\; \gamma^{ij}\in L^{\infty}(\mathbf{K}),
\end{equation}
where $D$ represents coordinate derivatives in any {\it normal}
coordinate system. With these hypotheses, if $n=3$ we have the continuous
embedding
\begin{equation}
 \label{sobo}
 H^1(\mathbf{K})\subset L^6(\mathbf{K}).
\end{equation}

\section{Free Scalar Fields in the Aeons}

We investigate the propagation of a scalar field  in the Aeon
$\bar{\mathcal{M}}$, that obeys the Klein-Gordon equation with a
variable mass $\bar{m}$ that is a measurable function on $\bar{\mathcal{M}}$:
\begin{equation}
\label{kgbarbu}
 \left[
   {\Box}_{\bar{g}}+\frac{n-1}{4n} R_{\bar{g}}+\bar{m}^2\right]\bar{u}=\bar{f}\;\;in\;\;\ring{\bar{\mathcal{M}}}.
\end{equation}
By the Liouville transform (\ref{liou}), this equation is equivalent to the 
equation (\ref{KG}) in $\ring{\bar{I}}\times\mathbf{K}$ where
\begin{equation}
 \label{}
 \hat{I}:=[\tau_-,0),\;\;\check{I}:=(0,\tau_+],\;\; \ring{\hat{I}}:=(\tau_-,0),\;\;\ring{\check{I}}:=(0,\tau_+).
\end{equation}
Since $(\mathcal{M},g)$ satisfies (\ref{mani}), this
equation has the very simple form
\begin{equation}
 \label{eqbar}
 \left[\partial_{\tau}^2-\Delta_{\mathbf K}+\frac{n-1}{4n}R_{\gamma}+\bar{m}^2\bar{\Omega}^2\right]\bar{\varphi}=\bar{\Omega}^{\frac{n+3}{2}}\bar{f},\;\;(\tau,\mathrm{x})\in\ring{\bar{I}}\times\mathbf{K}.
\end{equation}
We suppose that the mass and the conformal factor satisfy
\begin{equation}
 \label{condi}
 \bar{m}^2\bar{\Omega}^2\in L^1_{loc}\left(\bar{I};L^{\infty}(\mathbf{K})\right).
\end{equation}
The global Cauchy problem is easily solved:
\begin{Proposition}
 We assume that (\ref{rb}) and (\ref{condi}) hold. Then given
 $\tau_0\in\bar{I}$, $s\in[0,1]$, $\bar{\varphi}_0\in H^s(\mathbf{K})$,
 $\bar{\varphi}_1\in H^{s-1}(\mathbf{K})$, $\bar{f}\in
 L^1_{loc}\left(\bar{I};H^{s-1}(\mathbf{K})\right)$, the equation (\ref{eqbar}) has a
 unique solution $\bar{\varphi}$ satisfying
\begin{equation}
 \label{linfi}
 \bar{\varphi}\in L^{\infty}_{loc}\left(\bar{I};H^s(\mathbf{K})\right),\;\;
 \partial_{\tau}\bar{\varphi}\in L^{\infty}_{loc}\left(\bar{I};H^{s-1}(\mathbf{K})\right),
\end{equation}
\begin{equation}
 \label{init}
 \bar{\varphi}(\tau_0)=\bar{\varphi}_0,\;\;\partial_{\tau}\bar{\varphi}(\tau_0)=\bar{\varphi}_1.
\end{equation}
Moreover we have
\begin{equation}
 \label{regul}
 \bar{\varphi}\in
 C^0\left(\bar{I};H^s(\mathbf{K})\right)\cap
 C^1\left(\bar{I};H^{s-1}(\mathbf{K})\right)
\end{equation}
and there exists $C>0$ such that any solution satisfies
\begin{equation}
 \label{ineger}
\begin{split}
 \Vert\bar{\varphi}(\tau)\Vert_{H^s(\mathbf{K})}+\Vert\partial_{\tau}\bar{\varphi}(\tau)\Vert_{H^{s-1}(\mathbf{K})}\leq
C&\left(
  \Vert\bar{\varphi}_0\Vert_{H^s(\mathbf{K})}+\Vert\bar{\varphi}_1\Vert_{H^{s-1}(\mathbf{K})}+\left\vert\int_{\tau_0}^{\tau}\left\Vert\bar{\Omega}^{\frac{n+3}{2}}(\sigma)\bar{f}(\sigma)\right\Vert_{H^{s-1}(\mathbf{K})}d\sigma\right\vert\right)\\
&\times\exp\left(\left\vert\int_{\tau_0}^{\tau}\left\Vert\bar{m}(\sigma)\bar{\Omega}(\sigma)\right\Vert_{L^{\infty}(\mathbf{K})}^2d\sigma\right\vert\right).
\end{split}
\end{equation}
If we assume that
\begin{equation}
 \label{str}
 \bar{m}^2\bar{\Omega}^2\in
  L^1\left(\bar{I};L^{\infty}(\mathbf{K})\right),
\end{equation}
\begin{equation}
 \label{strf}
 \bar{\Omega}^{\frac{n+3}{2}}\bar{f}\in  L^1\left(\bar{I};H^{s-1}(\mathbf{K})\right),
\end{equation}
then the following limits exist:
\begin{equation}
 \label{limom}
 \bar{\psi}_0:=\lim_{\tau\rightarrow 0}\bar{\varphi}(\tau)\;\;in\;\;H^s(\mathbf{K}),
\end{equation}
\begin{equation}
 \label{limun}
 \bar{\psi}_1:=\lim_{\tau\rightarrow 0}\partial_{\tau}\bar{\varphi}(\tau)\;\;in\;\;H^{s-1}(\mathbf{K}).
\end{equation}
Furthermore, (\ref{str}) and (\ref{strf})  assure that given $ \bar{\psi}_0\in
H^s(\mathbf{K})$, $ \bar{\psi}_1\in
H^{s-1}(\mathbf{K})$, there exists a unique solution $\bar{\varphi}$
of (\ref{eqbar}) satisfying (\ref{regul}), (\ref{limom}),
(\ref{limun}) and the map $(\bar{\varphi}_0,\bar{\varphi}_1)\mapsto
(\bar{\psi}_0,\bar{\psi}_1)$ is a homeomorphism on
$H^s(\mathbf{K})\times H^{s-1}(\mathbf{K})$.
 \label{propco}
\end{Proposition}

The initial conditions (\ref{init}) make sense thanks to the Strauss
theorem \cite{strauss}:  (\ref{linfi})
implies that $\bar{\varphi}$ is a weakly continuous function with
values in $H^s(\mathbf{K})$ and since
$\bar{m}^2\bar{\Omega}^2\bar{\varphi}\in
L^{1}_{loc}(\bar{I};L^2(\mathbf{K}))$, the equation (\ref{eqbar})
assures that
$\partial_{\tau}^2\bar{\varphi}\in
L^{1}_{loc}(\bar{I};H^{s-2}(\mathbf{K}))$ and then
  $\partial_{\tau}\bar{\varphi}$ is a weakly continuous function with
  values in $H^{s-1}(\mathbf{K})$.\\

{\it Proof.} Due to the weak regularity of the coefficients, this
proposition is not a straight consequence of the classic results on
the hyperbolic Cauchy problem, but we can establish the existence of
the solution using a
standard method. We introduce the operator
\begin{equation}
 \label{AAA}
 \mathcal{A}:=
\left(
\begin{array}{cc}
0&1\\
-\Delta_{\mathbf{K}}+1&0
\end{array}
\right),\;\;Dom(\mathcal{A})=H^{s}(\mathbf{K})\times H^{s-1}(\mathbf{K}),
\end{equation}
that is selfadjoint in $H^{s-1}(\mathbf{K})\times H^{s-2}(\mathbf{K})$ and
we solve the integral equation
\begin{equation}
 \label{integeq}
\begin{split}
 \left(
\begin{array}{c}
\bar{\varphi}(\tau)\\
\bar{\psi}(\tau)
\end{array}
\right)
=\mathcal{F}\left(
\begin{array}{c}
\bar{\varphi}\\
\bar{\psi}
\end{array}
\right)(\tau)
:=&e^{i(\tau-\tau_0)\mathcal{A}}\left(
\begin{array}{c}
\bar{\varphi}_0\\
\bar{\varphi}_1
\end{array}
\right)
+\int_{\tau_0}^{\tau}e^{i(\tau-\sigma)\mathcal{A}}
\left(
\begin{array}{c}
0\\
\bar{\Omega}^{\frac{n+3}{2}}(\sigma)\bar{f}(\sigma) \end{array}
\right)
d\sigma\\
&+\int_{\tau_0}^{\tau}e^{i(\tau-\sigma)\mathcal{A}}
\left(
\begin{array}{c}
0\\
\left[1-\frac{n-1}{4n}R_{\gamma}-\bar{m}^2(\sigma)\bar{\Omega}^2(\sigma)\right]\bar{\varphi}(\sigma)
\end{array}
\right)
d\sigma.
\end{split}
\end{equation}
Since $s\in[0,1]$, we have for any $\Phi,\Psi\in X_h:=C^0\left([\tau_0-h,\tau_0+h]; H^s(\mathbf{K})\times
  H^{s-1}(\mathbf{K})\right)$ and $h>0$,
$$
\left\Vert\mathcal{F}(\Phi)-\mathcal{F}(\Psi)\right\Vert_{X_h}\leq
\left[\left(1+\Vert R_{\gamma}\Vert_{L^{\infty}}\right)h+\int_{\tau_0-h}^{\tau_0+h}\Vert\bar{m}(\sigma)\bar{\Omega}(\sigma)\Vert_{L^{\infty}}d\sigma\right]\left\Vert\Phi-\Psi\right\Vert_{X_h}.
$$
Therefore $\mathcal{F}$ is a contraction on $X_h$ for $h$ small
enough, and its unique fixed point $\bar{\Phi}:=(\bar{\varphi},\bar{\psi})$ is a
local solution of (\ref{integeq}). Moreover we have
\begin{equation*}
 \label{ingal}
\begin{split}
 \left\Vert
\left(
\begin{array}{c}
\bar{\varphi}(\tau)\\
\bar{\psi}(\tau)
\end{array}
\right)
\right\Vert_{H^s\times H^{s-1}}\leq
&\left\Vert
\left(
\begin{array}{c}
\bar{\varphi}_0\\
\bar{\varphi}_1
\end{array}
\right)
\right\Vert_{H^s\times H^{s-1}}
+\left\vert\int_{\tau_0}^{\tau}
\left\Vert
\bar{\Omega}^{\frac{n+3}{2}}(\sigma)\bar{f}(\sigma)\right\Vert_{H^{s-1}} d\sigma\right\vert\\
&+\left\vert\int_{\tau_0}^{\tau}\left(1+\Vert
    R_{\gamma}\Vert_{L^{\infty}}+\Vert\bar{m}^2(\sigma)\bar{\Omega}^2(\sigma)\Vert_{L^{\infty}}\right)
\Vert
\bar{\varphi}(\sigma)\Vert_{H^s} d\sigma\right\vert.
\end{split}
\end{equation*}
Therefore the Gronwall Lemma implies that
\begin{equation}
 \label{ingal}
\begin{split}
 \left\Vert
\left(
\begin{array}{c}
\bar{\varphi}(\tau)\\
\bar{\psi}(\tau)
\end{array}
\right)
\right\Vert_{H^s\times H^{s-1}}\leq
&C\left(\left\Vert
\left(
\begin{array}{c}
\bar{\varphi}_0\\
\bar{\varphi}_1
\end{array}
\right)
\right\Vert_{H^s\times H^{s-1}}
+\left\vert\int_{\tau_0}^{\tau}
\left\Vert
\bar{\Omega}^{\frac{n+3}{2}}(\sigma)\bar{f}(\sigma)\right\Vert_{H^{s-1}} d\sigma\right\vert\right)\\
&\times\exp\left(\left\vert\int_{\tau_0}^{\tau}\Vert\bar{m}^2(\sigma)\bar{\Omega}^2(\sigma)\Vert_{L^{\infty}} d\sigma\right\vert\right),
\end{split}
\end{equation}
and the principle
of the unique continuation assures that the local solution
$\bar{\Phi}$ can be
extended to a global solution $\bar{\Phi}\in
 C^0\left(\bar{I};H^s(\mathbf{K})\times H^{s-1}(\mathbf{K})\right)$ of the integral equation
 (\ref{integeq}) that obviously satisfies also  (\ref{ingal}) on
 $\bar{I}$. We warn  that in general $\bar{\Phi}\notin C^1\left( [\tau_0-h,\tau_0+h]; H^{s-1}\times
  H^{s-2}\right)$ since we do not assume $\bar{m}$ to be
continuous. Nevertheless $\bar{\psi}=\partial_{\tau}\bar{\varphi}$ and
$\bar{\varphi}$ is a global solution of (\ref{eqbar}) satisfying
(\ref{init}), (\ref{regul}) and (\ref{ineger}).
\\

To establish the
uniqueness, we recall (see
{\it e.g.} \cite{strauss}) that any $\varphi\in
L^{\infty}_{loc}(\bar{I};H^1(\mathbf{K}))$ with
$\partial_{\tau}\varphi\in L^{\infty}_{loc}(\bar{I};L^2(\mathbf{K}))$
solution in the sense of the distributions of
$(\partial^2_{\tau}-\Delta_{\mathbf{K}}+1)\varphi=F\in
L^1_{loc}(\bar{I};L^2(\mathbf{K}))$ satisfies
\begin{equation}
 \label{zegal}
 \Vert\varphi(\tau)\Vert^2_{H^1}+\Vert\partial_{\tau}\varphi(\tau)\Vert^2_{L^2}=
\Vert\varphi(\tau_0)\Vert^2_{H^1}+\Vert\partial_{\tau}\varphi(\tau_0)\Vert^2_{L^2}+2\Re\int_{\tau_0}^{\tau}\int_{\mathbf{K}}F(\sigma,\mathrm{x})\overline{\partial_{\tau}\varphi(\sigma,\mathrm{x})}d\mathrm{x} d\sigma,
\end{equation}
hence,
$$
\Vert\varphi(\tau)\Vert^2_{H^1}+\Vert\partial_{\tau}\varphi(\tau)\Vert^2_{L^2}\leq
\Vert\varphi(\tau_0)\Vert^2_{H^1}+\Vert\partial_{\tau}\varphi(\tau_0)\Vert^2_{L^2}+2\left\vert\int_{\tau_0}^{\tau}\Vert F(\sigma)\Vert_{L^2}\Vert\partial_{\tau}\varphi(\sigma)\Vert_{L^2} d\sigma\right\vert.
$$
We can apply this inequality to
$(-\Delta_{\mathbf{K}}+1)^{-\frac{1-s}{2}}\varphi$ where $\varphi\in
L^{\infty}_{loc} (\bar{I};H^s(\mathbf{K}))$ with
$\partial_{\tau}\varphi\in L^{\infty}_{loc} (\bar{I};H^{s-1}(\mathbf{K}))$ is
a solution in the sense of the distributions of
$(\partial^2_{\tau}-\Delta_{\mathbf{K}}+1)\varphi=F\in
L^1_{loc}(\bar{I};H^{s-1}(\mathbf{K}))$ to obtain
$$
 \Vert\varphi(\tau)\Vert^2_{H^s}+\Vert\partial_{\tau}\varphi(\tau)\Vert^2_{H^{s-1}}\leq
\Vert\varphi(\tau_0)\Vert^2_{H^s}+\Vert\partial_{\tau}\varphi(\tau_0)\Vert^2_{H^{s-1}}+2\left\vert\int_{\tau_0}^{\tau}\Vert F(\sigma)\Vert_{H^{s-1}}\Vert\partial_{\tau}\varphi(\sigma)\Vert_{H^{s-1}} d\sigma\right\vert.
$$
Taking
$F=\left(1-\frac{n-1}{4n}R_{\gamma}-\bar{m}^2\bar{\Omega}^2\right)\bar{\varphi}$
where $\bar{\varphi}$ is a solution of (\ref{eqbar}) with $\bar{f}=0$, 
this inequality implies
\begin{equation*}
 \begin{split}
 \Vert\bar{\varphi}(\tau)\Vert^2_{H^s}+\Vert\partial_{\tau}\bar{\varphi}(\tau)\Vert^2_{H^{s-1}}\leq &
\Vert\bar{\varphi}(\tau_0)\Vert^2_{H^s}+\Vert\partial_{\tau}\bar{\varphi}(\tau_0)\Vert^2_{H^{s-1}}\\
&+\left\vert\int_{\tau_0}^{\tau}\left(1+\Vert
    R_{\gamma}\Vert_{L^{\infty}}+\Vert\bar{m}^2(\sigma)\bar{\Omega}^2(\sigma)\Vert_{L^{\infty}}\right)\left(\Vert
    \bar{\varphi}(\sigma)\Vert_{H^{s}}^2+\Vert\partial_{\tau}\bar{\varphi}(\sigma)\Vert_{H^{s-1}}^2
  \right)d\sigma\right\vert
\end{split}
\end{equation*}
hence the uniqueness follows from the Gronwall lemma.\\

Now (\ref{ineger}), (\ref{str}) and (\ref{strf}) imply that $\bar{\varphi}\in
L^{\infty}(\bar{I};H^s(\mathbf{K}))$, then we have by (\ref{integeq})
\begin{equation}
 \label{limo}
\begin{split}
 \lim_{\tau\rightarrow 0}
\left(
\begin{array}{c}
\bar{\varphi}(\tau)\\
\partial_{\tau}\bar{\phi}(\tau)
\end{array}
\right)
=&e^{-i\tau_0\mathcal{A}}\left(
\begin{array}{c}
\bar{\varphi}_0\\
\bar{\varphi}_1
\end{array}
\right)
+\int_{\tau_0}^{0}e^{-i\sigma\mathcal{A}}
\left(
\begin{array}{c}
0\\
\bar{\Omega}^{\frac{n+3}{2}}(\sigma)\bar{f}(\sigma)
\end{array}
\right)
d\sigma\\
&+\int_{\tau_0}^{0}e^{-i\sigma\mathcal{A}}
\left(
\begin{array}{c}
0\\
\left[1-\frac{n-1}{4n}R_{\gamma}-\bar{m}^2(\sigma)\bar{\Omega}^2(\sigma)\right]\bar{\varphi}(\sigma)
\end{array}
\right)
d\sigma.
\end{split}
\end{equation}
Furthermore, assumptions (\ref{str}) and (\ref{strf}) allow to solve the initial value
problem with data given at $\tau=0$. Thanks to the integrability of
$\Vert\bar{m}\bar{\Omega}\Vert_{L^{\infty}}^2$, we  can mimick the previous proof
and solve the integral equation (\ref{integeq}) with
$\tau_0=0$. We conclude that for any $\tau_1,\tau_2\in\bar{I}\cup\{0\}$, the map
$\left(\bar{\varphi}(\tau_1),\partial_{\tau}\bar{\varphi}(\tau_1)\right)\mapsto
(\bar{\varphi}(\tau_2),\partial_{\tau}\bar{\varphi}(\tau_2))$ is a
homeomorphism on $H^s(\mathbf{K})\times H^{s-1}(\mathbf{K})$.

\fin

We immediately deduce the following:
\begin{Theorem}
We assume (\ref{rb}) and
\begin{equation}
 \label{condenor}
 \int_{\tau_-}^{\tau_+}\left(\left\Vert\tilde{\Omega}^{\frac{n+3}{2}}(\tau)\tilde{f}(\tau)\right\Vert_{H^{s-1}(\mathbf{K})}+\Vert\tilde{m}(\tau)\tilde{\Omega}(\tau)\Vert_{L^{\infty}(\mathbf{K})}^2\right)d\tau<\infty,
\end{equation}
where $\tilde{m}$, $\tilde{\Omega}$ and $\tilde{f}$ are defined by
(\ref{deftilde}). Then given $u_0\in H^s(\mathbf{K})$, $u_1\in
 H^{s-1}(\mathbf{K})$, $s\in[0,1]$, there exist unique solutions
 $\bar{u}\in C^0\left(\bar{I};H^s(\mathbf{K})\right)\cap
 C^1\left(\bar{I};H^{s-1}(\mathbf{K})\right)$ of (\ref{kgbarbu}) satisfying
\begin{equation}
 \label{initend}
 \hat{u}(\tau_-)=u_0,\;\;\partial_{\tau}\hat{u}(\tau_-)=u_1,
\end{equation}
\begin{equation}
 \label{trans}
 \lim_{\tau\rightarrow 0^+}\check{\Omega}(\tau)^{\frac{n-1}{2}}\check{u}(\tau)=\lim_{\tau\rightarrow 0^-}\hat{\Omega}^{\frac{n-1}{2}}(\tau)\hat{u}(\tau)\;\;in\;\;H^s(\mathbf{K}),
\end{equation}
\begin{equation}
 \label{transdt}
 \lim_{\tau\rightarrow 0^+}\partial_{\tau}\left[\check{\Omega}^{\frac{n-1}{2}}\check{u}\right](\tau)=\lim_{\tau\rightarrow 0^-}\partial_{\tau}\left[\hat{\Omega}^{\frac{n-1}{2}}\hat{u}\right](\tau)\;\;in\;\;H^{s-1}(\mathbf{K}).
\end{equation}
The function $\tilde{u}$ defined by
\begin{equation}
 \label{utilde}
 \tau\in[\tau_-,0)\Rightarrow
 \tilde{u}(\tau)=\hat{u}(\tau),\;\;\tau\in(0,\tau_+]\Rightarrow \tilde{u}(\tau)=\check{u}(\tau),
\end{equation}
satisfies
\begin{equation}
 \label{regtout}
\tilde{\Omega}^{\frac{n-1}{2}}\tilde{u}\in  C^0\left([\tau_-,\tau_+];H^s(\mathbf{K})\right)\cap
C^1\left([\tau_-,\tau_+];H^{s-1}(\mathbf{K})\right),
\end{equation}
\begin{equation}
 \label{eqtout}
 \left[\partial_{\tau}^2-\Delta_{\mathbf K}+\frac{n-1}{4n}R_{\gamma}+\tilde{m}^2\tilde{\Omega}^2\right]\left(\tilde{\Omega}^{\frac{n-1}{2}}\tilde{u}\right)=\tilde{\Omega}^{\frac{n+3}{2}}\tilde{f},\;\;(\tau,\mathrm{x})\in(\tau_-,\tau_+)\times\mathbf{K},
\end{equation}
The linear map $(\hat{u}(\tau_-),\partial_{\tau}\hat{u}(\tau_-))\mapsto (\check{u}(\tau_+),\partial_{\tau}\check{u}(\tau_+))$ is a
homeomorphism on $H^s(\mathbf{K})\times
H^{s-1}(\mathbf{K})$.
 \label{teofastoch}
\end{Theorem}

{\it Proof.}
We introduce the linear maps
\begin{equation}
 \label{LL}
 \hat{\mathfrak{L}}(\tau)=
\left(
  \begin{array}{cc}
    \hat{\Omega}^{\frac{n-1}{2}}(\tau)&0\\
    \partial_{\tau}\left[\hat{\Omega}^{\frac{n-1}{2}}\right](\tau)& \hat{\Omega}^{\frac{n-1}{2}}(\tau)
  \end{array}
\right),\;\;
\check{\mathfrak{L}}(\tau)=
\left(
  \begin{array}{cc}
    \check{\Omega}^{\frac{n-1}{2}}(\tau)&0\\
    \partial_{\tau}\left[\check{\Omega}^{\frac{n-1}{2}}\right](\tau)& \check{\Omega}^{\frac{n-1}{2}}(\tau)
  \end{array}
\right)
\end{equation}
that are isomorphisms on $H^s(\mathbf{K})\times
H^{s-1}(\mathbf{K})$. We apply the previous proposition. First we solve the
Cauchy problem (\ref{eqbar}), (\ref{init}) on $\hat{I}$ with
$(\hat{\varphi}_0,\hat{\varphi}_1)=\hat{\mathfrak{L}}(\tau_-)(u_0,u_1)$
and we put
$\hat{u}(\tau)=\hat{\Omega}^{\frac{1-n}{2}}(\tau)\hat{\varphi}(\tau)$. Then
we consider the solution $\check{\varphi}$ of (\ref{eqbar}) on $\check{I}$
satisfying $\check{\varphi}(0)=\hat{\psi}_0$,
$\partial_{\tau}\check{\varphi}(0)=\hat{\psi}_1$, and we put
$\check{u}(\tau)=\check{\Omega}^{\frac{1-n}{2}}(\tau)\check{\varphi}(\tau)$. Then
$\tilde{u}$ satisfies (\ref{initend}), (\ref{trans}), (\ref{transdt}),
and these transmission conditions imply that $\tilde{u}$ is solution
of (\ref{regtout}) and (\ref{eqtout}). Finally the maps
$(\hat{u}_0,\hat{u}_1)\mapsto
\hat{\mathfrak{L}}(\tau_-)(\hat{u}_0,\hat{u}_1)=(\hat{\varphi}_0,\hat{\varphi}_1)\mapsto(\hat{\psi}_0,\hat{\psi}_1)=(\check{\psi}_0,\check{\psi}_1)\mapsto(\check{\varphi}(\tau_+),\partial_{\tau}\check{\varphi}(\tau_+))\mapsto
\left[\check{\mathfrak{L}}(\tau_+)\right]^{-1}(\check{\varphi}(\tau_+),\partial_{\tau}\check{\varphi}(\tau_+))=(\check{u}(\tau_+),\partial_{\tau}\check{u}(\tau_+))$
are
homeomorphisms on $H^s(\mathbf{K})\times H^{s-1}(\mathbf{K})$.

\fin

This result allows to treat the situation of the Singular Bouncing Scenario
since the assumption  (\ref{condenor}) is very weak in this case: if
the source $\tilde{f}=0$, it
is sufficient that the mass is bounded and the convergences
(\ref{bounce}) are uniform.  Then we have a natural propagation from
the Previous Aeon to the Present Aeon despite the blow-up of the
fields $\hat{u}$ and $\check{u}$ at the Bang Surface, since the
normalized field $\tilde{\varphi}$ is a solution, continuous in time, of the equation
(\ref{KGbof}). In contrast  (\ref{condenor}) is a very
strong constraint for the expanding Aeons: for instance for the De Sitter like
metric (\ref{ds}) the mass has to decay exponentially so that
\begin{equation}
 \label{dskta}
 \int_{\tau_-}^0\hat{m}^2(\tau)\frac{d\tau}{\tau^2}\sim\int_{\hat{t}_-}^{\infty}\hat{m}^2(\hat{t})e^{\hat{H}\hat{t}}d\hat{t}<\infty.
\end{equation}
In the next section we relax this assumption to be able to treat the CCC
scenario with a much less drastic constraint on the decay of the mass.

\section{Asymptotics at the Bang Surface for a Slow Mass Decay}
In this part, we assume that the mass and the conformal factor depend
only on the time coordinate:
\begin{equation}
 \label{}
 \bar{m}\in C^0\left(\bar{I}\right),\;\;\bar{\Omega}\in C^2(\bar{I}).
\end{equation}
In this framework the hypothesis (\ref{condenor}) was
\begin{equation}
 \label{condenort}
 \int_{\bar{I}}\bar{m}^2(\tau)\bar{\Omega}^2(\tau)d\tau<\infty,
\end{equation}
and we now investigate the asymptotic behaviour of $\bar{\varphi}$, $\partial_{\tau}\bar{\varphi}$
under the weaker assumption
\begin{equation}
 \label{concon}
 \int_{\bar{I}}\bar{m}^2(\tau)\bar{\Omega}^2(\tau)\mid\tau \mid d\tau<\infty.
\end{equation}
In particular  for the De Sitter like
metric (\ref{ds}) this constraint is
\begin{equation}
 \label{dspakta}
 \int_{\tau_-}^0\hat{m}^2(\tau)\frac{d\tau}{\mid\tau\mid}\sim\int_{\hat{t}_-}^{\infty}\hat{m}^2(\hat{t})d\hat{t}<\infty
\end{equation}
that is much more weaker than (\ref{dskta}). Our strategy is based on
relating $\bar{m}^2\bar{\Omega}^2$ that, {\it a priori}, does not belong to
$L^1$, to an auxiliary fonction $\bar{A}$ that belongs to
$L^1$. These functions are linked by the Riccati equation
\begin{equation}
 \label{riccati}
 \bar{A}'-\bar{A}^2=\bar{m}^2\bar{\Omega}^2.
\end{equation}
This method has been initiated in \cite{delsanto} and used in \cite{RIP}. The main motivation of this approach is the following fundamental
result that treats the hard case of the blowing-up
$\partial_{\tau}\bar{\varphi}$ and describes its asymptotic behaviour
as $\tau\rightarrow 0$. To study the wave equation near $\tau=0$, we
introduce for $h>0$ small enough, the interval
\begin{equation}
 \label{}
 \bar{I}_h:=\bar{I}\cap [-h,h].
\end{equation}

\begin{Lemma}
 Assume there exist $h>0$ and $\bar{A}\in C^1\cap L^1(\bar{I}_h)$
 satisfying (\ref{riccati}) on $\bar{I}_h$.
Then there exists $C>0$ such that for the solution  $\bar{\varphi}\in
 C^0\left(\bar{I};H^s(\mathbf{K})\right)\cap
 C^1\left(\bar{I};H^{s-1}(\mathbf{K})\right)$ of (\ref{eqbar}),
 (\ref{init}) with $s\in[0,1]$ and $\bar{f}$ satisfying (\ref{strf}),
 we have for any $\tau\in\bar{I}_h$
 \begin{equation}
   \label{inegrat}
   \begin{split}
 \Vert\bar{\varphi}(\tau)\Vert_{H^s(\mathbf{K})}+
 \Vert\partial_{\tau}\bar{\varphi}(\tau)+&\bar{A}(\tau)\bar{\varphi}(\tau)\Vert_{H^{s-1}(\mathbf{K})}\leq\\
 C&\left(\Vert\bar{\varphi}_0\Vert_{H^s(\mathbf{K})}+\Vert\bar{\varphi}_1\Vert_{H^{s-1}(\mathbf{K})}+\left\vert\int_{\tau_0}^{\tau}\bar{\Omega}^{\frac{n+3}{2}}(\sigma)\Vert\bar{f}(\sigma)\Vert_{H^{s-1}(\mathbf{K})}d\sigma\right\vert\right),
 \end{split}
\end{equation}
and the
 following limits exist:
\begin{equation}
 \label{limpsio}
 \bar{\psi}_0:=\lim_{\tau\rightarrow 0}\bar{\varphi}(\tau)\;\;in\;\;H^s(\mathbf{K}),
\end{equation}
\begin{equation}
 \label{limpsiun}
 \bar{\psi}_1:=\lim_{\tau\rightarrow 0}\left(\partial_{\tau}\bar{\varphi}(\tau)+\bar{A}(\tau)\bar{\varphi}(\tau)\right)\;\;in\;\;H^{s-1}(\mathbf{K}).
\end{equation}
Moreover the map $W_{\bar{A}}:\;(\bar{\varphi}_0,\bar{\varphi}_1)\mapsto
(\bar{\psi}_0,\bar{\psi}_1)$ is a homeomorphism on
$H^s(\mathbf{K})\times H^{s-1}(\mathbf{K})$.
 \label{lemgaga}
\end{Lemma}

We emphasize that $\bar{A}(\tau)$ is allowed to blow-up as
$\tau\rightarrow 0$.\\

{\it Proof.}
Given $\bar{\tau}_h\in\bar{I}_h$, we put
\begin{equation}
 \label{psexp}
\bar{ \psi}(\bar{\tau}_h;\tau):=\bar{\varphi}(\tau)\exp\left(\int_{\bar{\tau}_h}^{\tau}\bar{A}(\sigma)d\sigma\right).
\end{equation}
Then $\bar{\psi}(\bar{\tau}_h;.)$ belongs to $C^0\left(\bar{I}_h;H^s(\mathbf{K})\right)\cap
 C^1\left(\bar{I}_h;H^{s-1}(\mathbf{K})\right)$ and is solution of 
\begin{equation}
 \label{eqpsi}
 \left(\partial^2_{\tau}-\Delta_{\mathbf{K}}+\frac{n-1}{4n}R_{\gamma}-2\bar{A}\partial_{\tau}\right)\bar{\psi}=\bar{F},\;\;\bar{F}:=\bar{\Omega}^{\frac{n+3}{2}}\bar{f}e^{\int_{\bar{\tau}_h}^{\tau}\bar{A}(\sigma)d\sigma},
\end{equation}
\begin{equation}
 \label{}
 \bar{\psi}(\bar{\tau}_h;\bar{\tau}_h)=\bar{\varphi}(\bar{\tau}_h),\;\;\partial_{\tau}\bar{\psi}(\bar{\tau}_h;\bar{\tau}_h)=\partial_{\tau}\bar{\varphi}(\bar{\tau}_h)+\bar{A}(\bar{\tau}_h)\bar{\varphi}(\bar{\tau}_h).
\end{equation}
Following the proof of the Proposition \ref{propco} , $\bar{\psi}(\bar{\tau}_h;.)$ is
the solution to the integral equation
\begin{equation}
  \label{integalspi}
  \begin{split}
 \left(
\begin{array}{c}
\bar{\psi}(\tau)\\
\partial_{\tau}\bar{\psi}(\tau)
\end{array}
\right)
=&e^{i(\tau-\bar{\tau}_h)\mathcal{A}}\left(
\begin{array}{c}
\bar{\psi}(\bar{\tau}_h;\bar{\tau}_h)\\
\partial_{\tau}\bar{\psi}(\bar{\tau}_h;\bar{\tau}_h)
\end{array}
\right)\\
&+\int_{\bar{\tau}_h}^{\tau}e^{i(\tau-\sigma)\mathcal{A}}
\left(
\begin{array}{c}
0\\
\left[1-\frac{n-1}{4n}R_{\gamma}\right]\bar{\psi}(\sigma)+2\bar{A}(\sigma)\partial_{\tau}\bar{\psi}(\sigma)+\bar{F}(\sigma)
\end{array}
\right)
d\sigma,
\end{split}
\end{equation}
and we have the energy inequality
\begin{equation*}
  \begin{split}
 \left\Vert
\left(
\begin{array}{c}
\bar{\psi}(\tau)\\
\partial_{\tau}\bar{\psi}(\tau)
\end{array}
\right)
\right\Vert_{H^s\times H^{s-1}}\leq
&\left\Vert
\left(
\begin{array}{c}
\bar{\psi}(\bar{\tau}_h;\bar{\tau}_h)\\
\partial_{\tau}\bar{\psi}(\bar{\tau}_h;\bar{\tau}_h)
\end{array}
\right)
\right\Vert_{H^s\times H^{s-1}}
+\left\vert\int_{\bar{\tau}_h}^{\tau}\Vert\bar{F}(\sigma)\Vert_{H^{s-1}}d\sigma\right\vert\\
&+\left\vert\int_{\bar{\tau}_h}^{\tau}\left(1+\Vert
    R_{\gamma}\Vert_{L^{\infty}}+2\vert \bar{A}(\sigma)\vert\right)
\left\Vert
\left(
\begin{array}{c}
\bar{\psi}(\sigma)\\
\partial_{\tau}\bar{\psi}(\sigma)
\end{array}
\right)
\right\Vert_{H^s\times H^{s-1}}
d\sigma\right\vert.
\end{split}
\end{equation*}
Since $\bar{A}\in L^1(\bar{I}_h)$,  the Gronwall Lemma assures
that there exists $C>0$ independent of $\tau$, $\bar{\tau}_h$ such that
\begin{equation}
 \left\Vert
\left(
\begin{array}{c}
\bar{\psi}(\tau)\\
\partial_{\tau}\bar{\psi}(\tau)
\end{array}
\right)
\right\Vert_{H^s\times H^{s-1}}\leq C\left(
\left\Vert
\left(
\begin{array}{c}
\bar{\psi}(\bar{\tau}_h;\bar{\tau}_h)\\
\partial_{\tau}\bar{\psi}(\bar{\tau}_h;\bar{\tau}_h)
\end{array}
\right)
\right\Vert_{H^s\times H^{s-1}}
+\left\vert\int_{\bar{\tau}_h}^{\tau}\bar{\Omega}^{\frac{n+3}{2}}(\sigma)\Vert\bar{f}(\sigma)\Vert_{H^{s-1}}d\sigma\right\vert\right).
\label{enerpsi}
\end{equation}
Using (\ref{ineger}) we have the estimate
\begin{equation}
  \label{tagada}
\left(
\begin{array}{c}
\bar{\psi}(\bar{\tau}_h;\bar{\tau}_h)\\
\partial_{\tau}\bar{\psi}(\bar{\tau}_h;\bar{\tau}_h)
\end{array}
\right)\leq C'
\left(
\left\Vert
\left(
\begin{array}{c}
\bar{\varphi}_0\\
\bar{\varphi}_1
\end{array}
\right)
\right\Vert_{H^s\times H^{s-1}}
+\left\vert\int_{\tau_0}^{\bar{\tau}_h}\bar{\Omega}^{\frac{n+3}{2}}(\sigma)\Vert\bar{f}(\sigma)\Vert_{H^{s-1}}d\sigma\right\vert\right),
\end{equation}
hence (\ref{inegrat}) is a consequence of (\ref{enerpsi}) and (\ref{tagada}).

Using the
integrability of $\bar{A}$ again, we can do $\tau\rightarrow 0$ in
(\ref{integalspi}) and we obtain the existence of the limits
(\ref{limpsio}), (\ref{limpsiun}) by putting
\begin{equation}
 \label{limingalspi}
 \left(
\begin{array}{c}
\bar{\psi}_0\\
\bar{\psi}_1
\end{array}
\right)
:=e^{-i\bar{\tau}_h\mathcal{A}}\left(
\begin{array}{c}
\bar{\psi}(\bar{\tau}_h;\bar{\tau}_h)\\
\partial_{\tau}\bar{\psi}(\bar{\tau}_h;\bar{\tau}_h)
\end{array}
\right)
+\int_{\bar{\tau}_h}^0e^{-i\sigma\mathcal{A}}
\left(
\begin{array}{c}
0\\
\left[1-\frac{n-1}{4n}R_{\gamma}\right]\bar{\psi}(\sigma)+2\bar{A}(\sigma)\partial_{\tau}\bar{\psi}(\sigma)+\bar{F}(\sigma)
\end{array}
\right)
d\sigma,
\end{equation}
moreover (\ref{enerpsi}) holds with $\tau=0$, hence $W_{\bar{A}}$ is a
well defined
continuous map on $H^s\times H^{s-1}$.

Conversely, given $(\bar{\psi}_0,\bar{\psi}_1)\in H^s\times H^{s-1}$, 
we can solve, as for Proposition \ref{propco}, the integral equation 
\begin{equation*}
 \label{ingalspi}
 \left(
\begin{array}{c}
\bar{\psi}(\tau)\\
\partial_{\tau}\bar{\psi}(\tau)
\end{array}
\right)
=e^{i\tau\mathcal{A}}\left(
\begin{array}{c}
\bar{\psi}_0\\
\partial_{\tau}\bar{\psi}_1
\end{array}
\right)
+\int_0^{\tau}e^{i(\tau-\sigma)\mathcal{A}}
\left(
\begin{array}{c}
0\\
\left[1-\frac{n-1}{4n}R_{\gamma}\right]\bar{\psi}(\sigma)+2\bar{A}(\sigma)\partial_{\tau}\bar{\psi}(\sigma)+\bar{F}(\sigma)
\end{array}
\right)
d\sigma
\end{equation*}
and
we obtain a unique solution $\bar{\psi}(0;.)\in C^0\left(\bar{I}_h\cup\{0\};H^s(\mathbf{K})\right)\cap
C^1\left(\bar{I}_h\cup\{0\};H^{s-1}(\mathbf{K})\right)$, and we have
\begin{equation}
  \label{tagadatsoin}
\left(
\begin{array}{c}
\bar{\psi}(0;\tau)\\
\partial_{\tau}\bar{\psi}(0;\tau)
\end{array}
\right)\leq C'
\left(
\left\Vert
\left(
\begin{array}{c}
\bar{\psi}_0\\
\bar{\psi}_1
\end{array}
\right)
\right\Vert_{H^s\times H^{s-1}}
+\left\vert\int_{0}^{\tau}\bar{\Omega}^{\frac{n+3}{2}}(\sigma)\Vert\bar{f}(\sigma)\Vert_{H^{s-1}}d\sigma\right\vert\right).
\end{equation}
Finally we
 solve the Cauchy problem for (\ref{eqbar}) with initial data
 $\bar{\varphi}(\bar{\tau}_h)=\bar{\psi}(0;\bar{\tau}_h)$,
 $\partial_{\tau}\bar{\varphi}(\bar{\tau}_h)=\partial_{\tau}\bar{\psi}(0;\bar{\tau}_h)-\bar{A}(\bar{\tau}_h)\bar{\psi}(0;\bar{\tau}_h)$,
 and we can invert $W_{\bar{A}}$ by putting $\bar{\varphi}_0:=\bar{\varphi}(\tau_0)$,
 $\bar{\varphi}_1:=\partial_{\tau}\bar{\varphi}(\tau_0)$. The
 bi-continuity of $W_{\bar{A}}:\;
 (\bar{\varphi}_0,\bar{\varphi}_1)\mapsto
 (\bar{\psi}_0,\bar{\psi}_1)$ is assured by the estimates
 (\ref{tagada}) and \ref{tagadatsoin}).

\fin

\begin{Lemma}
 Assume there exist $h>0$, $\hat{A}\in C^1\cap L^1([-h,0))$ and $\check{A}\in C^1\cap L^1((0,h])$
 satisfying (\ref{riccati}). Then given $\hat{\varphi}_0\in
 H^s(\mathbf{K})$, $\hat{\varphi}_1\in H^{s-1}(\mathbf{K})$ and $\bar{f}$ satisfying (\ref{strf}), there exists a
 unique $\tilde{\varphi}\in C^0\left([\tau_-,\tau_+];
   H^s(\mathbf{K})\right)\cap
 C^1\left([\tau_-,0)\cup(0,\tau_+];H^{s-1}(\mathbf{K})\right)$ such that
\begin{equation}
 \label{kiliki}
 \tilde{\varphi}(\tau_-)=\hat{\varphi}_0,\;\;\partial_{\tau}\tilde{\varphi}(\tau_-)=\hat{\varphi}_1,
\end{equation}
\begin{equation}
 \label{kileq}
 \left[\partial_{\tau}^2-\Delta_{\mathbf K}+\frac{n-1}{4n}R_{\gamma}+\tilde{m}^2\tilde{\Omega}^2\right]\tilde{\varphi}=\tilde{\Omega}^{\frac{n+3}{2}}\tilde{f},\;\;(\tau,\mathrm{x})\in(\tau_-,\tau_+) \setminus\{0\}\times\mathbf{K}
\end{equation}
and the function $\tilde{\psi}$ defined by
\begin{equation}
 \label{psipipi}
 \tilde{\psi}(\tau):=\tilde{\varphi}(\tau)\exp\left(\int_0^{\tau}\tilde{A}(\sigma)d\sigma\right)
\end{equation}
belongs to $C^0\left([-h,h];H^{s}(\mathbf{K})\right)\cap
 C^1\left([-h,h];H^{s-1}(\mathbf{K})\right),$ and satisfies
\begin{equation}
 \label{eqpopi}
 \left(\partial^2_{\tau}-\Delta_{\mathbf{K}}+\frac{n-1}{4n}R_{\gamma}-2\tilde{A}\partial_{\tau}\right)\tilde{\psi}=\tilde{\Omega}^{\frac{n+3}{2}}\tilde{f}
 e^{\int_0^{\tau}\tilde{A}(\sigma)d\sigma}\;\;in\;\;(-h,h)\times\mathbf{K},
\end{equation}
where $\tilde{m}$, $\tilde{\Omega}$ and $\tilde{f}$ are defined by (\ref{deftilde})
and $\tilde{A}\in L^1(-h,h)$ is given by 
\begin{equation}
 \label{atilde}
 \tau\in(-h,0)\Rightarrow \tilde{A}(\tau)=\hat{A}(\tau),\;\; \tau\in(0,h)\Rightarrow \tilde{A}(\tau)=\check{A}(\tau).
\end{equation}

The map
$S_{\tilde{A}}:\;\left(\tilde{\varphi}(\tau_-),\partial_{\tau}\tilde{\varphi}(\tau_-)\right)\mapsto
\left(\tilde{\varphi}(\tau_+),\partial_{\tau}\tilde{\varphi}(\tau_+)\right)$ is a
homeomorphism of $H^s(\mathbf{K})\times H^{s-1}(\mathbf{K})$.
 \label{lemkronk}
\end{Lemma}

{\it Proof.} For $\tau\in\hat{I}$ we define
$\tilde{\varphi}(\tau,.)=\hat{\varphi}(\tau,.)$ where $\hat{\varphi}$
is the solution on $\hat{I}$ of (\ref{eqbar}),
 (\ref{init}) with $\tau_0=\tau_-$. Using the notations of Lemma
 \ref{lemgaga}, we introduce
 $\left(\check{\varphi}_0,\check{\varphi}_1\right):=W_{\check{A}}^{-1}W_{\hat{A}}(\hat{\varphi}_0,\hat{\varphi}_1)$,
 and for $\tau\in\check{I}$ we define $\tilde{\varphi}(\tau,.)=\check{\varphi}(\tau,.)$ where $\check{\varphi}$
is the solution on $\check{I}$ of (\ref{eqbar}),
 (\ref{init}) with $\tau_0=\tau_+$. Since $\lim_{\tau\rightarrow
   0^-}\hat{\varphi}(\tau)=\lim_{\tau\rightarrow
   0^+}\check{\varphi}(\tau)$, $\tilde{\varphi}$ belongs to $ C^0\left([\tau_-,\tau_+];
   H^s(\mathbf{K})\right)\cap
 C^1\left([\tau_-,0)\cup(0,\tau_+];H^{s-1}(\mathbf{K})\right)$ and
it is the solution of (\ref{kiliki}) and (\ref{kileq}). Moreover $\tilde{\psi}$
defined by (\ref{psipipi}) satisfies
\begin{equation*}
 \label{}
 \tau\in (-h,0)\Rightarrow\hat{\psi}(\tau)=\hat{\psi}(\hat{\tau}_h;\tau)\exp\left(\int_0^{\hat{\tau}_h}\hat{A}(\sigma)d\sigma\right),
\;
 \tau\in (0,h)\Rightarrow\check{\psi}(\tau)=\check{\psi}(\check{\tau}_h;\tau)\exp\left(\int_0^{\check{\tau}_h}\check{A}(\sigma)d\sigma\right),
\end{equation*}
\begin{equation*}
 \label{}
 \lim_{\tau\rightarrow
   0^-}\hat{\psi}(\tau)=\lim_{\tau\rightarrow
   0^+}\check{\psi}(\tau),\;\;
  \lim_{\tau\rightarrow
   0^-}\partial_{\tau}\hat{\psi}(\tau)=\lim_{\tau\rightarrow
   0^+}\partial_{\tau}\check{\psi}(\tau),
\end{equation*}
therefore $\tilde{\psi}\in C^0\left([-h,h];H^{s}(\mathbf{K})\right)\cap
 C^1\left([-h,h];H^{s-1}(\mathbf{K})\right),$ and satisfies
(\ref{eqpopi}). Finally $S_{\tilde{A}}=W_{\check{A}}^{-1}W_{\hat{A}}$
is a homeomorphism.
\fin

Now given the mass and the conformal factor, we have to solve the
Riccati equation in $L^1$ near $\tau=0$. We need the
notation
\begin{equation}
 \label{etabar}
 \hat{\eta}:=+,\;\;\check{\eta}:=-.
\end{equation}
\begin{Proposition}

1) If $\bar{m}$ and $\bar{\Omega}$ satisfy (\ref{condenort}), then
given $\bar{\alpha}\in\RR,$ there exist
$h>0$ and a unique $\bar{A}\in
C^0(\{0\}\cup\bar{I}_h)\cap
C^1(\bar{I}_h)$ solution of (\ref{riccati}) with
$\bar{A}(0)=\bar{\alpha}$.\\

2) We assume that $\bar{m}$ and $\bar{\Omega}$ satisfy (\ref{concon}). Then for any $\epsilon>0$, there exist
$h_{\epsilon}>0$, $\tau_{\epsilon}\in\bar{I}_{h_{\epsilon}}$ and $\bar{A}_{\epsilon}\in
C^1(\bar{I}_{h_{\epsilon}})$ a real solution de (\ref{riccati}) satisfying
\begin{equation}
 \label{apslun}
 \bar{A}_{\epsilon}\in L^1(\bar{I}_{h_{\epsilon}}),
\end{equation}
\begin{equation}
 \label{inegala}
 \bar{\eta}\int_{\tau_{\epsilon}}^{\tau}\bar{m}^2(\sigma)\bar{\Omega}^2(\sigma)
 d\sigma\leq \bar{\eta}\bar{A}_{\epsilon}(\tau)\leq \bar{\eta}(1+\epsilon) \int_{\tau_{\epsilon}}^{\tau}\bar{m}^2(\sigma)\bar{\Omega}^2(\sigma)  d\sigma,
\end{equation}
\begin{equation}
 \label{integlop}
\bar{\eta}\int_{\tau_{\epsilon}}^0\mid\bar{A}_{\epsilon}(\tau)\mid d\tau
\leq \frac{\epsilon}{1+\epsilon}.
\end{equation}
The solutions $\bar{A}\in C^1(\bar{I}_h)$ of (\ref{riccati}) for some $h>0$ are,
either integrable, or satisfy
\begin{equation}
 \label{similo}
 \bar{A}(\tau)-\bar{A}_{\epsilon}(\tau)\sim
 -\tau^{-1},\;\;\tau\rightarrow 0.
\end{equation}
Moreover if $(\bar{m},\bar{\Omega})$  do
not satisfy (\ref{condenort}), then any real solution $\bar{A}\in
C^1\cap L^1(\bar{I}_h)$ of (\ref{riccati}) for some $h>0$, satisfies
\begin{equation}
 \label{boum}
 \hat{A}(\tau)\longrightarrow+\infty,\;\; \check{A}(\tau)\longrightarrow-\infty,\;\;\tau\rightarrow 0,
\end{equation}
\begin{equation}
 \label{difflim}
 \exists \bar{\alpha}:=\lim_{\tau\rightarrow 0}\bar{A}(\tau)-\bar{A}_{\epsilon}(\tau)\in\RR.
\end{equation}
 Furthermore, for any $\bar{\alpha}\in\RR$ there exists a unique $\bar{A}\in
L^1\cap
C^1(\bar{I}_h)$ solution of (\ref{riccati}) for some $h>0$, satisfying
(\ref{difflim}).
 \label{propric}
\end{Proposition}



{\it Proof.} 1) Since $\bar{m}\bar{\Omega}$ can be unbounded as
$\tau\rightarrow 0$, the first assertion is not a direct consequence
of the Cauchy-Lipschitz theorem but it can be proved by the usual way. If
$\bar{m}\bar{\Omega}\in L^2(\bar{I})$, we have to solve
\begin{equation}
 \label{integGAGA}
 \bar{A}(\tau)=\mathcal{G}(\bar{A})(\tau):=\bar{\alpha}+\int_0^{\tau}\bar{m}^2(\sigma)\bar{\Omega}^2(\sigma) d\sigma+\int_0^{\tau}\bar{A}^2(\sigma)d\sigma.
\end{equation}
We take $h>0$ small enough to that
$$
\int_{\bar{I}\cap[-h,h]}\bar{m}^2(\sigma)\bar{\Omega}^2(\sigma)
  d\sigma\leq 1,\;\;0<h<\frac{1}{\left(2+\mid\bar{\alpha}\mid\right)^2}.
$$
We can easily check that $\mathcal{G}$ is a strict contraction on
$\left\{A\in C^0(\{0\}\cup\bar{I}_h),\;\;\Vert
  A-\bar{\alpha}\Vert_{\infty}\leq 2\right\}$. Therefore its unique fixed point
$\bar{A}$ that is solution of (\ref{integGAGA}), satisfies
$\bar{A}(0)=\bar{\alpha}$ and the Riccati equation.\\

2) We first  construct $\hat{A}_{\epsilon}\in
C^1([\tau_{\epsilon},0))$ solution of
\begin{equation}
 \label{zedo}
 \hat{A}_{\epsilon}(\tau)=\int_{\tau_{\epsilon}}^{\tau}\bar{m}^2(\sigma)\bar{\Omega}^2(\sigma)d\sigma+\int_{\tau_{\epsilon}}^{\tau}\hat{A}^2_{\epsilon}(\sigma)d\sigma
\end{equation}
Thanks to the Fubini
theorem and (\ref{concon}) we choose
$\tau_{\epsilon}=-h_{\epsilon}\in(\tau_-,0)$ such that
\begin{equation}
 \label{fufu}
 \int_{\tau_{\epsilon}}^0\left(\int_{\tau_{\epsilon}}^{\tau}\bar{m}^2(\sigma)\bar{\Omega}^2(\sigma) d\sigma\right)d\tau=\int_{\tau_{\epsilon}}^0 \bar{m}^2(\tau)\bar{\Omega}^2(\tau)\mid\tau\mid d\tau\leq\frac{\epsilon}{(1+\epsilon)^2}.
\end{equation}
We define a sequence $A_n\in C^1([\tau_{\epsilon},0))$, $n\in\NN$,  by
\begin{equation}
 \label{secua}
 \forall\tau\in[\tau_{\epsilon},0),\;\;A_0(\tau)=0,\;\;A_{n+1}(\tau):=\int_{\tau_{\epsilon}}^{\tau}
\bar{m}^2(\sigma)\bar{\Omega}^2(\sigma)d\sigma+\int_{\tau_{\epsilon}}^{\tau}A_n^2(\sigma)d\sigma.
\end{equation}
We have $A_n\geq 0$ and
$$
A_1-A_0\geq 0,\;\;A_{n+1}(\tau)-A_n(\tau)=\int_{\tau_{\epsilon}}^{\tau}\left(A_n(\sigma)+A_{n-1}(\sigma)\right)\left((A_n(\sigma)-A_{n-1}(\sigma)\right)d\sigma,
$$
hence by recurrence we deduce that
\begin{equation}
 \label{croi}
 \int_{\tau_{\epsilon}}^{\tau}\bar{m}^2(\sigma)\bar{\Omega}^2(\sigma)d\sigma\leq
A_n(\tau)\leq A_{n+1}(\tau),\;\;n\geq 1.
\end{equation}
Assume that
\begin{equation}
 \label{recu}
 A_n(\tau)\leq (1+\epsilon) \int_{\tau_{\epsilon}}^{\tau}\bar{m}^2(\sigma)\bar{\Omega}^2(\sigma)d\sigma.
\end{equation}
Then we have with (\ref{fufu}):
\begin{equation*}
\begin{split}
 A_{n+1}(\tau)&\leq\int_{\tau_{\epsilon}}^{\tau}
\bar{m}^2(\sigma)\bar{\Omega}^2(\sigma)d\sigma+(1+\epsilon)^2\int_{\tau_{\epsilon}}^{\tau}\left(\int_{\tau_{\epsilon}}^{\sigma}\bar{m}^2(\zeta)\bar{\Omega}^2(\zeta)d\zeta\right)^2d\sigma\\
&\leq\int_{\tau_{\epsilon}}^{\tau}
\bar{m}^2(\sigma)\bar{\Omega}^2(\sigma)d\sigma
+(1+\epsilon)^2\int_{\tau_{\epsilon}}^{\tau}
\bar{m}^2(\sigma)\bar{\Omega}^2(\sigma)d\sigma\int_{\tau_{\epsilon}}^{\tau}\left(\int_{\tau_{\epsilon}}^{\sigma}\bar{m}^2(\zeta)\bar{\Omega}^2(\zeta)d\zeta\right)d\sigma\\
&\leq  (1+\epsilon) \int_{\tau_{\epsilon}}^{\tau}\bar{m}^2(\sigma)\bar{\Omega}^2(\sigma)d\sigma,
\end{split}
\end{equation*}
hence (\ref{recu}) is established for any $n$. Taking advantage of
(\ref{croi}) and (\ref{recu}) we can introduce the measurable function
\begin{equation}
 \label{}
 \hat{A}_{\epsilon}(\tau):=\lim_{n\rightarrow\infty}A_n(\tau)
\end{equation}
That satisfies (\ref{inegala}).
Thanks to the Beppo Levi theorem we deduce from (\ref{secua}) that
$\hat{A}_{\epsilon}$ satisfies also (\ref{zedo}) and then
$\hat{A}_{\epsilon}\in C^1([\tau_{\epsilon},0))$. Now (\ref{apslun}) and
(\ref{integlop}) follow from (\ref{inegala}) and (\ref{fufu}) by
integration. The construction of $\check{A}_{\epsilon}$ is similar: we
apply the previous procedure to $-\check{A}(-\tau)$.\\

Now given a real solution $\bar{A}\in C^1(\bar{I}_h))$
  of (\ref{riccati}), we have
$$
-h\leq\tau<0\Rightarrow\hat{A}(\tau)\geq A(-h)+\int_{-h}^{\tau}\bar{m}^2\bar{\Omega}^2d\sigma,\;\;
0<\tau\leq h\Rightarrow \check{A}(\tau)\leq \check{A}(h)+\int_h^{\tau}\bar{m}^2(\sigma)\bar{\Omega}^2(\sigma)d\sigma,
$$
hence (\ref{boum}) is satisfied if $(\bar{m},\bar{\Omega})$ do not
satisfy (\ref{condenort}). Now $\delta:=\bar{A}-\bar{A}_{\epsilon}$ is
solution of the equation
$$
\delta'(\tau)=\delta(\tau)\left[\bar{A}(\tau)+\bar{A}_{\epsilon}(\tau)\right],
$$
hence given $\tau_0\in\bar{I}_h\cap\bar{I}_{h_{\epsilon}}$ we have
$$
\delta(\tau)=\delta(\tau_0)\exp\left(\int_{\tau_0}^{\tau}\bar{A}(\sigma)+\bar{A}_{\epsilon}(\sigma)d\sigma\right)\longrightarrow\bar{\alpha}:=\delta(\tau_0)\exp\left(\int_{\tau_0}^0\bar{A}(\sigma)+\bar{A}_{\epsilon}(\sigma)d\sigma\right),\;\;\tau\rightarrow 0,
$$
and (\ref{difflim}) is proved. Moreover $\delta$ is also a solution of
the Bernouilli equation
$$
\delta'(\tau)-2\bar{A}_{\epsilon}(\tau)\delta(\tau)=\delta^2(\tau),
$$
hence if $\bar{A}\neq\bar{A}_{\epsilon}$, $\delta^{-1}$ is a solution of the linear equation
$$
\zeta'(\tau)+2\bar{A}_{\epsilon}(\tau)\zeta(\tau)=-1,
$$
then we conclude that all the real solutions
$\bar{A}\neq\bar{A}_{\epsilon}$ of the Riccati equation are given near $\tau=0$ by
\begin{equation}
 \label{}
 \bar{A}(\tau)=\bar{A}_{\epsilon}(\tau)+e^{2\int_{\tau_0}^{\tau}\bar{A}_{\epsilon}(\sigma)d\sigma}\left(\delta(\tau_0)-\int_{\tau_0}^{\tau}e^{2\int_{\tau_0}^{\sigma}\bar{A}_{\epsilon}(s)ds}d\sigma\right)^{-1},
\end{equation}
and these solutions are in $L^1$ near zero iff $\delta(\tau_0)\neq
\int_{\tau_0}^0e^{2\int_{\tau_0}^{\sigma}\bar{A}_{\epsilon}(s)ds}d\sigma$
and satisfy (\ref{similo}) if  $\delta(\tau_0)=
\int_{\tau_0}^0e^{2\int_{\tau_0}^{\sigma}\bar{A}_{\epsilon}(s)ds}d\sigma$.
Finally given $\bar{\alpha}\in\RR$, we put
\begin{equation}
 \label{}
 \bar{A}(\tau):=\bar{A}_{\epsilon}(\tau)+\bar{\alpha}e^{2\int_{0}^{\tau}\bar{A}_{\epsilon}(\sigma)d\sigma}\left(1+\bar{\alpha}\int_{\tau}^0e^{2\int_{0}^{\sigma}\bar{A}_{\epsilon}(s)ds}d\sigma\right)^{-1}
\end{equation}
that is a solution of (\ref{riccati}) that is well defined in $C^1\cap
L^1$ near
$\tau=0$ and satisfies (\ref{difflim}). To prove the uniqueness, we
consider two solutions $\bar{A},\bar{A}_*\in C^1\cap L^1(\bar{I}_h)$
satisfying (\ref{difflim}). Then given $\tau_0\in\bar{I}_h$ we have
$$
\left(\bar{A}(\tau_0)-\bar{A}_*(\tau_0)\right)exp\left(\int_{\tau_0}^{\tau}\bar{A}(\sigma)+\bar{A}_*(\sigma)d\sigma\right)\rightarrow
0,\;\;\tau\rightarrow 0.
$$
We conclude that $\bar{A}(\tau_0)=\bar{A}_*(\tau_0)$ and therefore $\bar{A}=\bar{A}_*$.
\fin

Since the existence of the $L^1$-solutions of the Riccati equation is
established, we can state the main result of this part.
\begin{Theorem}
  \label{teodur}
We assume (\ref{rb}) and
\begin{equation}
 \label{condenorgag}
 \int_{\tau_-}^0\hat{m}^2(\tau)\hat{\Omega}^2(\tau)\mid\tau\mid
 d\tau+\int_0^{\tau_+}\check{m}^2(\tau)\check{\Omega}^2(\tau) \tau d\tau<\infty.
\end{equation}
 Then given $u_0\in H^s(\mathbf{K})$, $u_1\in
 H^{s-1}(\mathbf{K})$, $s\in[0,1]$, and $\bar{f}$ satisfying (\ref{strf}),
given real solutions $\bar{A}\in C^1\cap L^1(\bar{I}_h))$
  of (\ref{riccati}) for some $h>0$,
there exist unique solutions
 $\bar{u}\in C^0\left(\bar{I};H^s(\mathbf{K})\right)\cap
 C^1\left(\bar{I};H^{s-1}(\mathbf{K})\right)$ of (\ref{kgbarbu}) satisfying
\begin{equation}
 \label{initendgag}
 \hat{u}(\tau_-)=u_0,\;\;\partial_{\tau}\hat{u}(\tau_-)=u_1,
\end{equation}
\begin{equation}
 \label{transgag}
 \lim_{\tau\rightarrow 0^+}\check{\Omega}^{\frac{n-1}{2}}(\tau)\check{u}(\tau)=\lim_{\tau\rightarrow 0^-}\hat{\Omega}^{\frac{n-1}{2}}(\tau)\hat{u}(\tau)\;\;in\;\;H^s(\mathbf{K}),
\end{equation}
\begin{equation}
 \label{transdtgag}
\begin{split}
 \lim_{\tau\rightarrow 0^+}&\left(\partial_{\tau}\left[\check{\Omega}^{\frac{n-1}{2}}\check{u}\right](\tau)+\check{A}(\tau)\check{\Omega}^{\frac{n-1}{2}}(\tau)\check{u}(\tau)\right)\\
&=\lim_{\tau\rightarrow 0^-}\left(\partial_{\tau}\left[\hat{\Omega}^{\frac{n-1}{2}}\hat{u}\right](\tau)+\hat{A}(\tau)\hat{\Omega}^{\frac{n-1}{2}}(\tau)\hat{u}(\tau)\right)
\;\;in\;\;H^{s-1}(\mathbf{K}).
\end{split}
\end{equation}
With the notations (\ref{deftilde}), (\ref{utilde}), (\ref{atilde}), we
have
\begin{equation}
  \label{}
  \tilde{\Omega}^{\frac{n-1}{2}}e^{\int_0^{\tau}\tilde{A}(\sigma)d\sigma}\tilde{u}
    \in
 C^0\left([-h,h];H^{s}(\mathbf{K})\right)\cap
 C^1\left([-h,h];H^{s-1}(\mathbf{K})\right),
\end{equation}
\begin{equation}
 \label{ekcefifi}
\left(\partial^2_{\tau}-\Delta_{\mathbf{K}}+\frac{n-1}{4n}R_{\gamma}-2\tilde{A}\partial_{\tau}\right)\left(\tilde{\Omega}^{\frac{n-1}{2}}e^{\int_0^{\tau}\tilde{A}(\sigma)d\sigma}\tilde{u}\right)=\tilde{\Omega}^{\frac{n+3}{2}}e^{\int_0^{\tau}\tilde{A}(\sigma)d\sigma}\tilde{f}\;\;in\;\;(-h,h)\times\mathbf{K},
\end{equation}

The linear map $\mathfrak{S}:\;(\hat{u}(\tau_-),\partial_{\tau}\hat{u}(\tau_-))\mapsto (\check{u}(\tau_+),\partial_{\tau}\check{u}(\tau_+))$ is a
continuous homeomorphism on $H^s(\mathbf{K})\times H^{s-1}(\mathbf{K})$.
 \label{}
\end{Theorem}

\begin{Remark}
  \label{}
  If (\ref{str}) is satisfied, the transmissions conditions (\ref{trans}),
  (\ref{transdt}) correspond to (\ref{transgag}), (\ref{transdtgag}) by
choosing the solutions $\bar{A}$ of the Riccati equation with $\bar{A}(0)=0$. There is less real freedom in  (\ref{transdtgag}) than is apparent from the two
  arbitrary functions $\hat{A}$ and $\check{A}$. In fact these
  transmission conditions (\ref{transgag}), (\ref{transdtgag}) form
  a one-real-parameter family: if we fix two solutions
  $\hat{A}_{\epsilon}$, $\check{A}_{\epsilon}$, then (\ref{difflim})
  and (\ref{transgag})
  assure that putting
  $\delta:=\hat{\alpha}-\check{\alpha}$, (\ref{transdtgag})
  is equivalent to
  \begin{equation}
 \label{transdtgaga}
\begin{split}
 \lim_{\tau\rightarrow 0^+}&\left(\partial_{\tau}\left[\check{\Omega}^{\frac{n-1}{2}}\check{u}\right](\tau)+\check{A}_{\epsilon}(\tau)\check{\Omega}^{\frac{n-1}{2}}(\tau)\check{u}(\tau)\right)\\
&=\lim_{\tau\rightarrow 0^-}\left(\partial_{\tau}\left[\hat{\Omega}^{\frac{n-1}{2}}\hat{u}\right](\tau)+\left(\hat{A}_{\epsilon}(\tau)+\delta\right)\hat{\Omega}^{\frac{n-1}{2}}(\tau)\hat{u}(\tau)\right)
\;\;in\;\;H^{s-1}(\mathbf{K}).
\end{split}
\end{equation}
Conversely, given $\delta\in\RR$, the last assertion of Proposition
\ref{propric} assures that there exists $\bar{A}$ satisfying
(\ref{transgag}). We conclude that the whole family of the conditions
(\ref{transgag}), (\ref{transdtgag}) indexed by the integrable
solutions $\bar{A}$ of the Riccati equation, is reduced to the
one-parameter family of transmission conditions (\ref{transgag}),
(\ref{transdtgaga}) indexed by the real parameter $\delta\in\RR$.

\end{Remark}

{\it Proof.}
The existence of $\hat{u}$ is given by Proposition
\ref{propco} by taking
$\hat{u}(\tau):=\hat{\Omega}^{\frac{1-n}{2}}(\tau)\hat{\varphi}(\tau)$
where $\hat{\varphi}$ is the solution of (\ref{eqbar}), (\ref{init})  with the
initial data
$(\hat{\varphi}_0,\hat{\varphi}_1)=\hat{\mathfrak{L}}(\tau_-)(u_0,u_1)$ given
at $\tau_0=\tau_-$ and $\hat{\mathfrak{L}}$ is defined by (\ref{LL}). Lemma \ref{lemgaga} assures that the following limits
exist
$$
\hat{\psi}_0:=\lim_{\tau\rightarrow
  0^-}\hat{\Omega}^{\frac{n-1}{2}}(\tau)\hat{u}(\tau)\;\;in\;\;H^s(\mathbf{K}),
$$
$$
\hat{\psi}_1:=\lim_{\tau\rightarrow 0^-}\left(\partial_{\tau}\left[\hat{\Omega}^{\frac{n-1}{2}}\hat{u}\right](\tau)+\hat{A}(\tau)\hat{\Omega}^{\frac{n-1}{2}}(\tau)\hat{u}(\tau)\right)\;\;in\;\;H^{s-1}(\mathbf{K}),
$$
and we can define
$$
(\check{\varphi}_0,\check{\varphi}_1):=W_{\check{A}}^{-1}(\hat{\psi}_0,\hat{\psi}_1).
$$
We now define
$\check{u}(\tau):=\check{\Omega}^{\frac{1-n}{2}}(\tau)\check{\varphi}(\tau)$
with the solution $\check{\varphi}$ of of (\ref{eqbar}), (\ref{init})
with $\tau_0=\tau_+$. Then (\ref{transgag}) and (\ref{transdtgag}) are direct
consequences of this construction. Finally we invoke Lemma \ref{lemkronk} to
conclude that (\ref{ekcefifi}) is deduced from (\ref{eqpopi}) and
$$
\mathfrak{S}=\left[\check{\mathfrak{L}}(\tau_+)\right]^{-1}S_{\tilde{A}}\hat{\mathfrak{L}}(\tau_-)
$$
is a homeomorphism on $H^s(\mathbf{K})\times H^{s-1}(\mathbf{K})$.

\fin

It is well known that the Riccati equations cannot be solved by
quadrature. Nevertheless we can give an explicit formulation of the
transmission conditions in the following important case.
\begin{Example} We consider the case where
  \begin{equation}
 \label{}
 \bar{m}^2(\tau)\bar{\Omega}^2(\tau)=\bar{c}^2\mid\tau\mid^{-1}+\bar{F}(\tau)
\end{equation}
  with $\bar{c}>0$ and $\bar{F}$ is a holomorphic function on a neigborhood
  of zero. We take a (generalized) eigenfunction $\Phi_{\lambda}\in
  L^2_{loc}(\mathbf{K})$ solution of $\left(-\Delta_{\mathbf
      K}+\frac{n-1}{4n}R_{\gamma}\right)\Phi_{\lambda}=\lambda\Phi_{\lambda}$, $\lambda\in\RR$.
  The solutions $\bar{\varphi}$ of (\ref{eqbar}) with $\bar{f}=0$, of type
  $\bar{\varphi}(\tau,\mathrm{x})=\bar{\phi}(\tau)\Phi_{\lambda}(\mathrm{x})$,
  are defined by the solutions $\bar{\phi}$ of
    the ODE
\begin{equation}
 \label{fuchs}
 \bar{\phi}''(\tau)+\left(\lambda+\frac{\bar{c}^2}{\mid
  \tau\mid}+\bar{F}(\tau)\right)\bar{\phi}(\tau)=0,\;\;\tau\in\bar{I}.
\end{equation}
We can solve this ODE by the Frobenius method (see {\it e.g.} Theorem
4.5 in
\cite{teschl}). Since $\tau=0$ is a
regular singular point, the Fuchs theorem assures that there exist two
functions $\bar{h}_1$, $\bar{h}_2$ which are analytic near zero with
$\bar{h}_1(0)=\bar{h}_2(0)=1$, and the general solution of
(\ref{fuchs}) can be written as
\begin{equation}
 \label{}
 \bar{\phi}(\tau)=\bar{C}_1\tau\bar{h}_1(\tau)+\bar{C}_2\left[\bar{h}_2(\tau)-\bar{c}^2\mid\tau \mid\bar{h}_1(\tau)\ln(\mid\tau\mid)\right],\;\;\bar{C}_j\in\CC.
\end{equation}
We deduce that
\begin{equation}
 \label{}
 \lim_{\tau\rightarrow
   0}\bar{\phi}(\tau)=\bar{C}_2,\;\;\bar{\phi}'(\tau)=
 \bar{\eta}\bar{C}_2\bar{c}^2\ln(\mid\tau\mid)+\bar{C}_1+\bar{C}_2\left(\bar{h}_2'(0)+\bar{\eta}\bar{c}^2\right)+o(1),\;\tau\rightarrow 0,
\end{equation}
with $\bar{\eta}$ defined by (\ref{etabar}).
Now the Riccati equation (\ref{riccati}) is reduced to a linear second
order ODE by the usual way, by putting
$\bar{A}=-\frac{\bar{\alpha}'}{\bar{\alpha}}$ where $\bar{\alpha}$ is
a solution of 
\begin{equation*}
 \label{}
 \bar{\alpha}''(\tau)+\left(\frac{\bar{c}^2}{\mid  \tau\mid}+\bar{F}(\tau)\right)\bar{\alpha}(\tau)=0.
\end{equation*}
We apply the Fuchs theorem again: there exist two
functions $\bar{k}_1$, $\bar{k}_2$, holomorphic near zero, with $\bar{k}_j(0)=1$ and 
$$
\bar{\alpha}(\tau)=\bar{D}_1\tau\bar{k}_1(\tau)+\bar{D}_2\left[\bar{k}_2(\tau)-\bar{c}^2\mid\tau \mid\bar{k}_1(\tau)\ln(\mid\tau\mid)\right],\;\;\bar{D}_j\in\CC.
$$
We get that the solutions of the Riccati equation are given by
\begin{equation*}
 \label{}
 \bar{A}(\tau)=-\frac{\bar{D}_1\left[\bar{k}_1(\tau)+\tau\bar{k}'_1(\tau)\right]+\bar{D}_2\left[\bar{k}'_2(\tau)+\bar{\eta}\bar{c}^2\bar{k}_1(\tau)\ln(\mid\tau\mid)-\bar{c}^2\mid\tau \mid\bar{k}'_1(\tau)\ln(\mid\tau\mid)+\bar{\eta}\bar{c}^2\bar{k}_1(\tau)\right]}{\bar{D}_1\tau\bar{k}_1(\tau)+\bar{D}_2\left[\bar{k}_2(\tau)-\bar{c}^2\mid\tau\mid \bar{k}_1(\tau)\ln(\mid\tau\mid)\right]}.
\end{equation*}
If $\bar{D}_2=0$ we have $\bar{A}(\tau)\sim -\tau^{-1}$ as
$\tau\rightarrow 0$. We conclude that the solutions of the Riccati
equation that are integrable near zero satisfy $\bar{D}_2\neq 0$, and
then
\begin{equation*}
 \label{}
 \bar{A}(\tau)=
 -\bar{\eta}\bar{c}^2\ln(\mid\tau\mid)-\frac{\bar{D}_1}{\bar{D}_2}-\bar{k}'_2(0)-\bar{\eta}\bar{c}^2+o(1),\;\;\tau\rightarrow 0.
\end{equation*}
Then putting
\begin{equation*}
 \label{}
 \delta:=\hat{h}'_2(0)-\check{h}'_2(0)+\check{k}'_2(0)-\hat{k}'_2(0)
 +\frac{\check{D}_1}{\check{D}_2}-\frac{\hat{D}_1}{\hat{D}_2},
\end{equation*}
the family of transmission conditions (\ref{transgag}), (\ref{transdtgag}) has the form
\begin{equation}
 \label{}
\check{C}_2= \hat{C}_2,
\end{equation}
\begin{equation}
 \label{}
\check{C}_1= \hat{C}_1+\delta\hat{C}_2
\end{equation}
where the real parameter $\delta$ can be arbitrarily choosen in $\RR$.
\end{Example}

The previous example is the unique case of choice of
$\bar{m}\bar{\Omega}$ for which  the
Klein-Gordon equation (\ref{eqbar}), expressed as a first order system, is equivalent to a Fuchsian system
$$U'(\tau)=AU(\tau)+\tau^{-1}B(\tau) U(\tau)+C(\tau)$$ where $B$ is analytic
near zero. In this simple linear situation, we could use the Frobenius
method. We mention that for non linear Fuchsian systems
$$U'(\tau)=AU(\tau)+\tau^{-1}B\left(\tau,U(\tau)\right)+C(\tau)$$
arising in General Relativity, another specific Fuchsian analysis has been developped (see {\it e.g.} \cite{beyer2010},
  \cite{beyer2019}, \cite{fournodavlos}, \cite{rendall}).

\section{self-interacting scalar field}
In this section we assume that the complete manifold $\mathbf{K}$ is a 3-dimensional
$C^{\infty}$ {\it
  bounded geometry manifold}, and we
investigate the massive semilinear Klein-Gordon equation
\begin{equation}
\label{nlkg}
 \left[
   {\Box}_{\bar{g}}+\frac{1}{6}
   R_{\bar{g}}+\bar{m}^2\right]\bar{u}=-\kappa\mid \bar{u}\mid^2\bar{u}\;\;in\;\;\bar{\mathcal{M}},
\end{equation}
where $\kappa>0$ is a coupling constant. The Liouville transform
$\bar{\varphi}:=\bar{\Omega}\bar{u}$ leads to the equivalent equation
\begin{equation}
 \label{nlKG}
 \left[
   {\Box}_{g}+\frac{1}{6}R_{\gamma}+\bar{m}^2\bar{\Omega}^2\right]\bar{\varphi}=-\kappa\mid
 \bar{\varphi}\mid^2\bar{\varphi}\;\;\;in\;\;\bar{\mathcal{M}}.
\end{equation}

First we suppose that the mass and the conformal factor satisfy
(\ref{condi}) and we solve the global Cauchy problem.
\begin{Proposition}
 We assume that (\ref{condi}) holds. Then given
 $\tau_0\in\bar{I}$, $\bar{\varphi}_0\in H^1(\mathbf{K})$,
 $\bar{\varphi}_1\in L^2(\mathbf{K})$, the equation (\ref{nlKG}) has a
 unique solution $\bar{\varphi}$ satisfying
\begin{equation}
 \label{regulnl}
 \bar{\varphi}\in
 C^0\left(\bar{I};H^1(\mathbf{K})\right)\cap
 C^1\left(\bar{I};L^2(\mathbf{K})\right)
\end{equation}
\begin{equation}
 \label{initnl}
 \bar{\varphi}(\tau_0)=\bar{\varphi}_0,\;\;\partial_{\tau}\bar{\varphi}(\tau_0)=\bar{\varphi}_1.
\end{equation}
Moreover there exists $C>0$ such that any solution satisfies
\begin{equation}
  \label{inegernl}
  \begin{split}
 \Vert\bar{\varphi}(\tau)\Vert_{H^1(\mathbf{K})}+\Vert\partial_{\tau}\bar{\varphi}(\tau)\Vert_{L^2(\mathbf{K})}\leq
C&\left(
  \Vert\bar{\varphi}_0\Vert_{H^1(\mathbf{K})}+\Vert\bar{\varphi}_0\Vert^2_{H^1(\mathbf{K})}+\Vert\bar{\varphi}_1\Vert_{L^2(\mathbf{K})}\right)\\
&\times\exp\left(\frac{1}{2}\left\vert\int_{\tau_0}^{\tau}\left\Vert\bar{m}(\sigma)\bar{\Omega}(\sigma)\right\Vert_{L^{\infty}(\mathbf{K})}^2d\sigma\right\vert\right),
\end{split}
\end{equation}
and the map $(\bar{\varphi}_0,\bar{\varphi}_1)\in H^1\times
L^2\mapsto\bar{\varphi}\in C^0(\bar{I};H^1)\cap C^1(\bar{I};L^2)$ is
continuous.\\

If we assume (\ref{str}),
then the following limits exist:
\begin{equation}
 \label{limomnl}
 \bar{\psi}_0:=\lim_{\tau\rightarrow 0}\bar{\varphi}(\tau)\;\;in\;\;H^1(\mathbf{K}),
\end{equation}
\begin{equation}
 \label{limunnl}
 \bar{\psi}_1:=\lim_{\tau\rightarrow 0}\partial_{\tau}\bar{\varphi}(\tau)\;\;in\;\;L^2(\mathbf{K}).
\end{equation}
Furthermore, (\ref{str}) assures that given $ \bar{\psi}_0\in
H^1(\mathbf{K})$, $ \bar{\psi}_1\in
L^2(\mathbf{K})$, there exists a unique solution $\bar{\varphi}$
of (\ref{nlKG}) satisfying (\ref{regulnl}), (\ref{limomnl}),
(\ref{limunnl}) and the map $(\bar{\varphi}_0,\bar{\varphi}_1)\mapsto
(\bar{\psi}_0,\bar{\psi}_1)$ is a bi-Lipschitz bijection on
$H^1(\mathbf{K})\times L^2(\mathbf{K})$.
 \label{propconl}
\end{Proposition}
{\it Proof.}
First we prove the local existence of the mild solutions by a classic
way. We solve the integral equation
\begin{equation}
  \label{integeqnl}
 \left(
\begin{array}{c}
\bar{\varphi}(\tau)\\
\bar{\psi}(\tau)
\end{array}
\right)
=\mathcal{F}\left(
\begin{array}{c}
\bar{\varphi}\\
\bar{\psi}
\end{array}
\right)(\tau),
\end{equation}
with
\begin{equation}
  \label{nlF}
 \mathcal{F}\left(
\begin{array}{c}
\bar{\varphi}\\
\bar{\psi}
\end{array}
\right)(\tau):=e^{i(\tau-\tau_0)\mathcal{A}}\left(
\begin{array}{c}
\bar{\varphi}_0\\
\bar{\varphi}_1
\end{array}
\right)
+\int_{\tau_0}^{\tau}e^{i(\tau-\sigma)\mathcal{A}}
\left(
\begin{array}{c}
0\\
\left[1-\frac{1}{6}R_{\gamma}-\bar{m}^2(\sigma)\bar{\Omega}^2(\sigma)-\kappa\mid\varphi(\sigma)\mid^2\right]\bar{\varphi}(\sigma)
\end{array}
\right)
d\sigma
\end{equation}
where $\mathcal{A}$ is defined by (\ref{AAA}).
Using the Sobolev inequality
\begin{equation}
 \label{sobo}
 \Vert u\Vert_{L^6(\mathbf{K})}\leq K \Vert
u\Vert_{H^1(\mathbf{K})},
\end{equation}
we get
\begin{equation*}
 \label{}
\begin{split}
 \left\Vert \mathcal{F}\left(
\begin{array}{c}
\bar{\varphi}\\
\bar{\psi}
\end{array}
\right)(\tau)\right\Vert_{H^1\times L^2}
\leq&
\left\Vert \left(
\begin{array}{c}
\bar{\varphi}_0\\
\bar{\varphi}_1
\end{array}
\right)\right\Vert_{H^1\times L^2}
+\left\vert\int_{\tau_0}^{\tau}\Vert\bar{m}(\sigma)\bar{\Omega}(\sigma)\Vert^2_{L^{\infty}}d\sigma\right\vert
\sup_{(\tau_0,\tau)}\Vert\bar{\varphi}(\sigma)\Vert_{H^1}\\
  &+\mid\tau-\tau_0\mid\left(1+\Vert
    R_{\gamma}\Vert_{L^{\infty}}+\kappa K^3
  \sup_{(\tau_0,\tau)}\Vert\bar{\varphi}(\sigma)\Vert^2_{H^1}\right)\sup_{(\tau_0,\tau)}\Vert\bar{\varphi}(\sigma)\Vert_{H^1}.
\end{split}
\end{equation*}
Putting
\begin{equation}
 \label{JJ}
 \rho:=\Vert(\bar{\varphi}_0,\bar{\phi}_1)\Vert_{H^1\times
  L^2},\;\;J_{\epsilon}:=\{\tau\in\bar{I};\;\;\mid\tau-\tau_0\mid\leq \epsilon\}\;\;(\epsilon>0),
\end{equation}
\begin{equation}
 \label{BB}
 B_{\rho}:=\left\{(\bar{\varphi},\bar{\psi})\in
C^0\left(J_{\epsilon};H^1(\mathbf{K})\times
  L^2(\mathbf{K})\right);\;\;\sup_{\tau\in J_{\epsilon}}\Vert
(\bar{\varphi}(\tau),\bar{\psi}(\tau))\Vert_{H^1\times L^2}\leq
2\rho\right\},
\end{equation}
we deduce that $\mathcal{F}$ is a map from $B_{\rho}$ into $B_{\rho}$
if
$$
2\int_{J_{\epsilon}}\Vert\bar{m}(\sigma)\bar{\Omega}(\sigma)\Vert^2_{L^{\infty}}d\sigma+\epsilon\left(1+\Vert
    R_{\gamma}\Vert_{L^{\infty}}+4\kappa K^3\rho^2\right)\leq 1.
$$
We have also:
\begin{equation*}
 \label{}
 \left\Vert \mathcal{F}\left(
\begin{array}{c}
\bar{\varphi}\\
\bar{\psi}
\end{array}
\right)(\tau)
-
\mathcal{F}\left(
\begin{array}{c}
\bar{\varphi}_*\\
\bar{\psi}_*
\end{array}
\right)(\tau)\right\Vert_{H^1\times L^2}
\leq \mathfrak{K}(\tau) \sup_{(\tau_0,\tau)}\Vert\bar{\varphi}(\sigma)-\bar{\varphi}_*(\sigma)\Vert_{H^1},
\end{equation*}
where
\begin{equation*}
\mathfrak{K}(\tau):=\left\vert\int_{\tau_0}^{\tau}\Vert\bar{m}(\sigma)\bar{\Omega}(\sigma)\Vert^2_{L^{\infty}}d\sigma\right\vert+\mid\tau-\tau_0\mid\left[1+\Vert
    R_{\gamma}\Vert_{L^{\infty}}+2\kappa K^3\left(
  \sup_{(\tau_0,\tau)}\Vert\bar{\varphi}(\sigma)\Vert^2_{H^1}+ \sup_{(\tau_0,\tau)}\Vert\bar{\varphi}_*(\sigma)\Vert^2_{H^1}\right)\right].
\end{equation*}
Hence $\mathcal{F}$ is a strict contraction on $B_{\rho} $ if
$$
2\int_{J_{\epsilon}}\Vert\bar{m}(\sigma)\bar{\Omega}(\sigma)\Vert^2_{L^{\infty}}d\sigma+\epsilon\left(1+\Vert
    R_{\gamma}\Vert_{L^{\infty}}+16\kappa K^3\rho^2\right)< 1.
  $$
  Its unique fixed point satisfies
  $\bar{\psi}=\partial_{\tau}\bar{\varphi}$ and it is a local solution of (\ref{nlKG}), (\ref{initnl})
  in $C^0\left(J_{\epsilon};H^1(\mathbf{K})\right)\cap C^1\left(J_{\epsilon};L^2(\mathbf{K})\right)$.
Now we deduce from (\ref{zegal}) that this local solution satisfies
\begin{equation}
 \label{inerrrnl}
\begin{split}
 \Vert\bar{\varphi}(\tau)\Vert^2_{H^1}+\Vert\partial_{\tau}\bar{\varphi}(\tau)\Vert^2_{L^2}+
\frac{\kappa}{2}\Vert\bar{\varphi}(\tau)\Vert_{L^4}^4\leq&
\Vert\bar{\varphi}_0\Vert^2_{H^1}+\Vert\bar{\varphi}_1\Vert^2_{L^2}+
\frac{\kappa}{2}\Vert\bar{\varphi}_0\Vert_{L^4}^4\\
&+\left\vert\int_{\tau_0}^{\tau}\left(1+\Vert
    R_{\gamma}\Vert_{L^{\infty}}+\Vert\bar{m}(\sigma)\bar{\Omega}(\sigma)\Vert^2_{L^{\infty}}\right)\left(\Vert\bar{\varphi}(\sigma)\Vert_{L^2}^2+\Vert\partial_{\tau}\bar{\varphi}(\sigma)\Vert^2_{L^2}\right)d\sigma\right\vert,
\end{split}
\end{equation}
hence we get by the Gronwall Lemma,
\begin{equation}
 \label{}
 \begin{split}
 \Vert\bar{\varphi}(\tau)\Vert^2_{H^1}+\Vert\partial_{\tau}\bar{\varphi}(\tau)\Vert^2_{L^2}+
\frac{\kappa}{2}\Vert\bar{\varphi}(\tau)\Vert_{L^4}^4\leq &
\left(\Vert\bar{\varphi}_0\Vert^2_{H^1}+\Vert\bar{\varphi}_1\Vert^2_{L^2}+
\frac{\kappa}{2}\Vert\bar{\varphi}_0\Vert_{L^4}^4\right)\\
&\times\exp{\left\vert\int_{\tau_0}^{\tau}\left(1+\Vert
    R_{\gamma}\Vert_{L^{\infty}}+\Vert\bar{m}(\sigma)\bar{\Omega}(\sigma)\Vert^2_{L^{\infty}}\right)d\sigma\right\vert},
\end{split}
\end{equation}
and (\ref{inegernl}) is proved for $\tau\in J_{\epsilon}$. Now the global
existence on the whole interval $\bar{I}$ follows from the principle
of unique continuation. We can easily prove the continuous dependence of
the solution with respect to the initial data. Given another solution
$\bar{\varphi}_*$ we have
\begin{equation*}
  \begin{split}
\left\Vert \left(
\begin{array}{c}
\bar{\varphi}\\
\partial_{\tau}\bar{\varphi}
\end{array}
\right)(\tau)
-
\left(
\begin{array}{c}
\bar{\varphi}_*\\
\partial_{\tau}\bar{\varphi}_*
\end{array}
\right)(\tau)\right\Vert_{H^1\times L^2}
\leq&
\left\Vert \left(
\begin{array}{c}
\bar{\varphi}\\
\partial_{\tau}\bar{\varphi}
\end{array}
\right)(\tau_0)
-
\left(
\begin{array}{c}
\bar{\varphi}_*\\
\partial_{\tau}\bar{\varphi}_*
\end{array}
\right)(\tau_0)\right\Vert_{H^1\times L^2}\\
&
+C\left\vert\int_{\tau_0}^{\tau}\Vert\bar{m}(\sigma)\bar{\Omega}(\sigma)\Vert_{L^{\infty}}^2
    \Vert\bar{\varphi}(\sigma)-\bar{\varphi}_*(\sigma)\Vert_{H^1}d\sigma\right\vert\\
&+C\left\vert\int_{\tau_0}^{\tau}\left(1+\Vert\bar{\varphi}(\sigma)\Vert_{H^1}^2+\Vert\bar{\varphi}_*(\sigma)\Vert_{H^1}^2\right)\Vert\bar{\varphi}(\sigma)-\bar{\varphi}_*(\sigma)\Vert_{H^1}d\sigma\right\vert
\end{split}
\end{equation*}
hence the Lipschitz property of the map
$(\bar{\varphi}_0,\bar{\varphi}_1)\mapsto\bar{\varphi}$ follows from
the Gronwall Lemma and (\ref{inegernl}).\\

Finally we assume that (\ref{str}) is satisfied. Then
(\ref{inegernl}) implies that
$\bar{\varphi}(\tau)$ is bounded in $H^1(\mathbf{K})$ hence we can
take the limit of the integral in (\ref{nlF}) as $\tau\rightarrow 0$,
and  we obtain (\ref{limomnl}), (\ref{limunnl}). Furthermore we can
take $\tau_0=0$ in (\ref{nlF}) and repeating the previous arguments, we
can solve the global Cauchy problem with initial data specified at
$\tau=0$. The proof is complete.
\fin

As a consequence we obtain directly  the following:
\begin{Theorem}
We assume (\ref{str}).
 Then given $u_0\in H^1(\mathbf{K})$, $u_1\in
 L^2(\mathbf{K})$, there exist unique solutions
 $\bar{u}\in C^0\left(\bar{I};H^1(\mathbf{K})\right)\cap
 C^1\left(\bar{I};L^2(\mathbf{K})\right)$ of (\ref{nlkg}) satisfying
\begin{equation}
 \label{initendnl}
 \hat{u}(\tau_-)=u_0,\;\;\partial_{\tau}\hat{u}(\tau_-)=u_1,
\end{equation}
\begin{equation}
 \label{transnl}
 \lim_{\tau\rightarrow 0^+}\check{\Omega}(\tau)\check{u}(\tau)=\lim_{\tau\rightarrow 0^-}\hat{\Omega}(\tau)\hat{u}(\tau)\;\;in\;\;H^1(\mathbf{K}),
\end{equation}
\begin{equation}
 \label{transdtnl}
 \lim_{\tau\rightarrow 0^+}\partial_{\tau}\left[\check{\Omega}\check{u}\right](\tau)=\lim_{\tau\rightarrow 0^-}\partial_{\tau}\left[\hat{\Omega}\hat{u}\right](\tau)\;\;in\;\;L^2(\mathbf{K}).
\end{equation}
Moreover $\tilde{u}$ defined by (\ref{utilde}) satisfies
\begin{equation}
 \label{regtoutun}
\tilde{\Omega}\tilde{u}\in  C^0\left([\tau_-,\tau_+];H^1(\mathbf{K})\right)\cap
C^1\left([\tau_-,\tau_+];L^2(\mathbf{K})\right),
\end{equation}
\begin{equation}
 \label{eqnlu}
 \left[
   {\Box}_{g}+\frac{1}{6}R_{\gamma}+\tilde{m}^2\tilde{\Omega}^2\right]\left[\tilde{\Omega}\tilde{u}\right]=-\kappa\tilde{\Omega}^3\mid
 \tilde{u}\mid^2\tilde{u}\;\;\;in\;\;\mathcal{M}.
\end{equation}
The map $(\hat{u}(\tau_-),\partial_{\tau}\hat{u}(\tau_-))\mapsto (\check{u}(\tau_+),\partial_{\tau}\check{u}(\tau_+))$ is a
bi-Lipschitz bijection on $H^1(\mathbf{K})\times L^2(\mathbf{K})$.
 \label{teofastochnl}
\end{Theorem}

{\it Proof.} We apply the previous proposition with $\tau_0=\tau_-$
and
$(\hat{\varphi}_0,\hat{\varphi}_1)=\hat{\mathfrak{L}}(\tau_-)(u_0,u_1)$,
and we put
$\hat{u}(\tau)=\hat{\Omega}^{-1}(\tau)\hat{\varphi}(\tau)$. Then we
consider the solution $\check{\varphi}$ of (\ref{nlKG}) on $\check{I}$
satisfying
$(\check{\psi}_0,\check{\psi}_1)=(\hat{\psi}_0,\hat{\psi}_1)$ and we
put
$\check{u}(\tau)=\check{\Omega}^{-1}(\tau)\check{\varphi}(\tau)$. Therefore
$\bar{u}$ are solutions of (\ref{nlkg}) and satisfy (\ref{transnl}),
(\ref{transdtnl}), and these transmission conditions imply that
$\tilde{u}$ satisfies also (\ref{regtoutun}) and
(\ref{eqnlu}). Finally the maps
$(u_0,u_1)\mapsto(\hat{\varphi}_0,\hat{\varphi}_1)\mapsto(\hat{\psi}_0,\hat{\psi}_1)=(\check{\psi}_0,\check{\psi}_1)\mapsto(\check{\varphi}(\tau_+),\partial_{\tau}\check{\varphi}(\tau_+))\mapsto(\check{u}(\tau_+),\partial_{\tau}\check{u}(\tau_+))$
are bi-Lipschitz bijections of $H^1(\mathbf{K})\times L^2(\mathbf{K})$.
\fin

Like for the linear case, this result is suitable to treat the case of
the Singular Bouncing Scenario but (\ref{str}) is a much too strong
assumption for the CCC. We now consider the more reasonnable
assumption (\ref{condenorgag}).

\begin{Theorem}
We assume
(\ref{condenorgag}).
 Then given $u_0\in H^1(\mathbf{K})$, $u_1\in
 L^2(\mathbf{K})$,
given real solutions $\bar{A}\in C^1\cap L^1(\bar{I}_h)$
  of (\ref{riccati}) for some $h>0$,
there exist unique solutions
 $\bar{u}\in C^0\left(\bar{I};H^1(\mathbf{K})\right)\cap
 C^1\left(\bar{I};L^2(\mathbf{K})\right)$ of (\ref{nlkg}) satisfying
\begin{equation}
 \label{initendgagnl}
 \hat{u}(\tau_-)=u_0,\;\;\partial_{\tau}\hat{u}(\tau_-)=u_1,
\end{equation}
\begin{equation}
 \label{transgagnl}
 \lim_{\tau\rightarrow 0^+}\check{\Omega}(\tau)\check{u}(\tau)=\lim_{\tau\rightarrow 0^-}\hat{\Omega}(\tau)\hat{u}(\tau)\;\;in\;\;H^1(\mathbf{K}),
\end{equation}
\begin{equation}
 \label{transdtgagnl}
\begin{split}
 \lim_{\tau\rightarrow 0^+}\left(\partial_{\tau}\left[\check{\Omega}\check{u}\right](\tau)+\check{A}(\tau)\check{\Omega}(\tau)\check{u}(\tau)\right)
=\lim_{\tau\rightarrow 0^-}\left(\partial_{\tau}\left[\hat{\Omega}\hat{u}\right](\tau)+\hat{A}(\tau)\hat{\Omega}(\tau)\hat{u}(\tau)\right)
\;\;in\;\;L^2(\mathbf{K}).
\end{split}
\end{equation}

With the notations (\ref{deftilde}), (\ref{utilde}), (\ref{atilde}), we
have
\begin{equation}
  \label{}
  \tilde{\psi}:=\tilde{\Omega}e^{\int_0^{\tau}\tilde{A}(\sigma)d\sigma}\tilde{u}
    \in
 C^0\left([-h,h];H^1(\mathbf{K})\right)\cap
 C^1\left([-h,h];L^2(\mathbf{K})\right),
\end{equation}
\begin{equation}
 \label{ekcefifika}
\left(\partial^2_{\tau}-\Delta_{\mathbf{K}}+\frac{1}{6}R_{\gamma}-2\tilde{A}\partial_{\tau}\right)\tilde{\psi}=-\kappa\mid\tilde{\psi}\mid^2\tilde{\psi}
e^{-2\int_0^{\tau}\tilde{A}(\sigma)d\sigma}\;\;in\;\;(-h,h)\times\mathbf{K},
\end{equation}

The map $\mathfrak{S}:\;(\hat{u}(\tau_-),\partial_{\tau}\hat{u}(\tau_-))\mapsto (\check{u}(\tau_+),\partial_{\tau}\check{u}(\tau_+))$ is a
bi-Lipschitz bijection on $H^1(\mathbf{K})\times L^2(\mathbf{K})$.
 \label{teodurnl}
\end{Theorem}


{\it Proof.} $\hat{u}$ is given by Proposition \ref{propconl} by
putting for $\tau\in\hat{I}$,
$\hat{u}(\tau)=\hat{\Omega}^{-1}(\tau)\hat{\varphi}(\tau)$ where
$\hat{\varphi}$ is the solution of (\ref{nlKG}), (\ref{initnl}) with
$\tau_0=\tau_-$,
$(\hat{\varphi}_0,\hat{\varphi}_1)=\hat{\mathfrak{L}}(\tau_-)(u_0,u_1)$. Now
for $\tau\in\hat{I}_h$ we
introduce
$$
\hat{\psi}(\tau):=e^{\int_0^{\tau}\hat{A}(\sigma)d\sigma}\hat{\varphi}(\tau)
$$
that is a solution of
$$
\left(\partial^2_{\tau}-\Delta_{\mathbf{K}}+1\right)\hat{\psi}=
\left(1+2\hat{A}\partial_{\tau}-\frac{1}{6}R_{\gamma}\right)\hat{\psi}
-\kappa\mid\hat{\psi}\mid^2\hat{\psi}
e^{-2\int_0^{\tau}\hat{A}(\sigma)d\sigma}\;\;in\;\;\hat{I}_h\times\mathbf{K}.
$$
Since $\hat{\psi}\in C^0(\hat{I}_h;H^1(\mathbf{K}))\cap
C^1(\hat{I}_h;L^2(\mathbf{K}))$, we deduce from (\ref{zegal}) that
\begin{equation*}
 \label{}
 \begin{split}
\Vert\hat{\psi}(\tau)\Vert_{H^1}^2+\Vert\partial_{\tau}\hat{\psi}(\tau)\Vert_{L^2}^2\leq&
C\left[\Vert u_0\Vert^2_{H^1}+\Vert
  u_1\Vert_{L^2}^2+\int_{\tau_-}^{\tau}(1+\mid\hat{A}(\sigma)\mid)\left(
    \Vert\hat{\psi}(\sigma)\Vert_{H^1}^2+\Vert\partial_{\sigma}\hat{\psi}(\sigma)\Vert_{L^2}^2\right)d\sigma\right]\\
&-\frac{\kappa}{2}\int_{\tau_-}^{\tau}\frac{d}{d\sigma}\left(\Vert\hat{\psi}(\sigma)\Vert_{L^4}^4\right)e^{-2\int_0^{\sigma}\hat{A}(s)ds}d\sigma,
\end{split}
\end{equation*}
hence
\begin{equation*}
 \label{}
 \begin{split}
\Vert\hat{\psi}(\tau)\Vert_{H^1}^2+\Vert\partial_{\tau}\hat{\psi}(\tau)\Vert_{L^2}^2+\frac{\kappa}{2}\Vert\hat{\psi}(\tau)\Vert_{L^4}^4e^{-2\int_0^{\tau}\hat{A}(s)ds}\leq&
C'\left[\Vert u_0\Vert^2_{H^1}+\Vert u_0\Vert^4_{H^1}+\Vert
  u_1\Vert_{L^2}^2\right]\\
+C''\int_{\tau_-}^{\tau}(1+\mid\hat{A}(\sigma)\mid)&\left(\Vert\hat{\psi}(\sigma)\Vert_{H^1}^2+\Vert\partial_{\sigma}\hat{\psi}(\sigma)\Vert_{L^2}^2+\frac{\kappa}{2}\Vert\hat{\psi}(\sigma)\Vert_{L^4}^4e^{-2\int_0^{\sigma}\hat{A}(s)ds}\right)d\sigma
 \end{split}
\end{equation*}
therefore we conclude with the Gronwall Lemma that
\begin{equation}
 \label{estiKRAK}
 \Vert\hat{\psi}(\tau)\Vert_{H^1}^2+\Vert\partial_{\tau}\hat{\psi}(\tau)\Vert_{L^2}^2\leq C'''\left[\Vert u_0\Vert^2_{H^1}+\Vert u_0\Vert^4_{H^1}+\Vert
  u_1\Vert_{L^2}^2\right]\exp\left(\int_{\tau_-}^{\tau}(1+\mid\hat{A}(\sigma)\mid)d\sigma\right)
\end{equation}
and finally since $\hat{A}\in L^1(\hat{I}_h)$,
\begin{equation}
 \label{bobopsi}
 \hat{\psi}\in L^{\infty}(\hat{I}_h;H^1(\mathbf{K})),\;\; \partial_{\tau}\hat{\psi}\in L^{\infty}(\hat{I}_h;L^2(\mathbf{K})).
\end{equation}
Now we have
\begin{equation*}
  \begin{split}
 \left(
   \begin{array}{c}
     \hat{\psi}(\tau)\\
     \partial_{\tau}\hat{\psi}(\tau)
   \end{array}
 \right)
 =&e^{i(\tau-\tau_-)\mathcal{A}}
 \left(
   \begin{array}{c}
     \hat{\psi}(\tau_-)\\
     \partial_{\tau}\hat{\psi}(\tau_-)
   \end{array}
 \right)\\
 +&
 \int_{\tau_-}^{\tau}e^{i(\tau-\sigma)\mathcal{A}}
 \left(
   \begin{array}{c}
     0\\
     \hat{\psi}(\sigma)-\frac{1}{6}R_{\gamma}\hat{\psi}(\sigma)+2\hat{A}(\sigma)\partial_{\sigma}\hat{\psi}(\sigma)
     -\kappa\mid\hat{\psi}(\sigma)\mid^2\hat{\psi}(\sigma)e^{-2\int_0^{\sigma}\hat{A}(s)ds}
   \end{array}
 \right)d\sigma
 \end{split}
\end{equation*}
We deduce from (\ref{bobopsi}) that the following limits exist
\begin{equation}
  \label{}
  \begin{split}
 \lim_{\tau\rightarrow
   0^-}
 &\left(
   \begin{array}{c}
     \hat{\Omega}(\tau)\hat{u}(\tau)\\
     \partial_{\tau}\left[\hat{\Omega}\hat{u}\right](\tau)+\hat{A}(\tau)\hat{u}(\tau)
   \end{array}
 \right)
 =
 \lim_{\tau\rightarrow
   0^-}\left(
   \begin{array}{c}
     \hat{\psi}(\tau)\\
     \partial_{\tau}\hat{\psi}(\tau)
   \end{array}
 \right)\\
 &=e^{-i\tau_-\mathcal{A}}
 \left(
   \begin{array}{c}
     \hat{\psi}(\tau_-)\\
     \partial_{\tau}\hat{\psi}(\tau_-)
   \end{array}
 \right)\\
 +
&
 \int_{\tau_-}^{0}e^{-i\sigma\mathcal{A}}
 \left(
   \begin{array}{c}
     0\\
     \hat{\psi}(\sigma)-\frac{1}{6}R_{\gamma}\hat{\psi}(\sigma)+2\hat{A}(\sigma)\partial_{\sigma}\hat{\psi}(\sigma)
     -\kappa\mid\hat{\psi}(\sigma)\mid^2\hat{\psi}(\sigma)e^{-2\int_0^{\sigma}\hat{A}(s)ds}
   \end{array}
 \right)d\sigma.
 \end{split}
\end{equation}
Now we define $\hat{\psi}_0:=\lim_{\tau\rightarrow
  0^-}\hat{\psi}(\tau)$,  $\hat{\psi}_1:=\lim_{\tau\rightarrow
  0^-}\partial_{\tau}\hat{\psi}(\tau)$, and we look for
$\check{\psi}\in C^0\left(\check{I}_h;H^1(\mathbf{K})\right)\cap
C^1\left(\check{I}_h;L^2(\mathbf{K})\right)$ solution
of
$$
\left(\partial^2_{\tau}-\Delta_{\mathbf{K}}+1\right)\check{\psi}=
\left(1+2\check{A}\partial_{\tau}-\frac{1}{6}R_{\gamma}\right)\check{\psi}
-\kappa\mid\check{\psi}\mid^2\check{\psi}
e^{-2\int_0^{\tau}\check{A}(\sigma)d\sigma}\;\;in\;\;\check{I}_h\times\mathbf{K},
$$
satisfying
$$
\lim_{\tau\rightarrow 0^+}\check{\psi}(\tau)=\hat{\psi}_0,\;\; \lim_{\tau\rightarrow 0^+}\partial_{\tau}\check{\psi}(\tau)=\hat{\psi}_1.
$$
It is sufficient to establish that the Cauchy problem is well posed in
$C^0\left((-h,h);H^1(\mathbf{K})\right)\cap
C^1\left((-h,h);L^2(\mathbf{K})\right)$ for equation (\ref{ekcefifika})
with an initial data $(\tilde{\psi}_0,\tilde{\psi}_1)$ given at any given time $\tau_0\in [-h,h]$. In
an equivalent way, we have to solve the integral equation
\begin{equation}
 \label{zekinkin}
 \begin{split}
 \left(
   \begin{array}{c}
     \tilde{\psi}(\tau)\\
     \tilde{\chi}(\tau)
   \end{array}
 \right)&=\mathcal{G} \left(
   \begin{array}{c}
     \tilde{\psi}\\
     \tilde{\chi}
   \end{array}
 \right)(\tau)\\
 &:=e^{i(\tau-\tau_0)\mathcal{A}}
 \left(
   \begin{array}{c}
     \tilde{\psi}_0\\
     \tilde{\psi}_1
   \end{array}
 \right)\\
 +&
 \int_{\tau_0}^{\tau}e^{i(\tau-\sigma)\mathcal{A}}
 \left(
   \begin{array}{c}
     0\\
     \tilde{\psi}(\sigma)-\frac{1}{6}R_{\gamma}\tilde{\psi}(\sigma)+2\tilde{A}(\sigma)\tilde{\chi}(\sigma)
     -\kappa\mid\tilde{\psi}(\sigma)\mid^2\tilde{\psi}(\sigma)e^{-2\int_0^{\sigma}\tilde{A}(s)ds}
   \end{array}
 \right)d\sigma.
 \end{split}
\end{equation}
We define
$$
\rho:=\Vert(\tilde{\psi}_0,\tilde{\psi}_1)\Vert_{H^1\times
  L^2},\;\;M:=\max\left(1+\frac{1}{6}\Vert R_{\gamma}\Vert_{L^{\infty}},\kappa\exp\left(2\Vert\tilde{A}\Vert_{L^1(-h,h)}\right)\right),
$$
then using (\ref{sobo})  and the notations (\ref{JJ}) and (\ref{BB}),
we get for any $(\tilde{\psi},\tilde{\chi})\in B_{\rho}$
$$
\sup_{\tau\in J_{\epsilon}}\left\Vert \mathcal{G} \left(
   \begin{array}{c}
     \tilde{\psi}\\
     \tilde{\chi}
   \end{array}
 \right)(\tau)\right\Vert_{H^1\times L^2}
 \leq
 \rho\left[1+2M(1+4K^3\rho^2)\epsilon+4\int_{J_{\epsilon}}\mid\tilde{A}(\tau)\mid d\tau\right],
$$
hence $\mathcal{G}$ is a map from $B_{\rho}$ into $B_{\rho}$ if
$\epsilon>0$ is small enough to that
$$
2M(1+4K^3\rho^2)\epsilon+4\int_{J_{\epsilon}}\mid\tilde{A}(\tau)\mid
d\tau \leq1.
$$
Moreover, given $(\tilde{\psi},\tilde{\chi}),
(\tilde{\psi}_*,\tilde{\chi}_*)\in B_{\rho}$, we have
\begin{equation*}
 \label{}
 \begin{split}
\sup_{\tau\in J_{\epsilon}}\left\Vert \mathcal{G} \left(
   \begin{array}{c}
     \tilde{\psi}\\
     \tilde{\chi}
   \end{array}
 \right)(\tau)
 -
 \mathcal{G} \left(
   \begin{array}{c}
     \tilde{\psi}_*\\
     \tilde{\chi}_*
   \end{array}
 \right)(\tau)
 \right\Vert_{H^1\times L^2}
 &\leq\epsilon M(1+12K^3\rho^2)\sup_{\tau\in
   J_{\epsilon}}\Vert\tilde{\psi}(\tau)-\tilde{\psi}_*(\tau)\Vert_{H^1}\\
 &+2\left(\int_{J_{\epsilon}}\mid\tilde{A}(\tau)\mid dt\right)\sup_{\tau\in
   J_{\epsilon}}\Vert\tilde{\chi}(\tau)-\tilde{\chi}_*(\tau)\Vert_{L^2}.
   \end{split}
 \end{equation*}
 We conclude that if
 $$
2M(1+6K^3\rho^2)\epsilon+4\int_{J_{\epsilon}}\mid\tilde{A}(\tau)\mid
d\tau <1,
$$
then $\mathcal{G}$ is a strict contraction on $B_{\rho}$ and its fixed
point satisfies $\tilde{\chi}=\partial_{\tau}\tilde{\psi}$ and
$\tilde{\psi}\in C^0\left(J_{\epsilon};H^1(\mathbf{K})\right)\cap
C^1\left(J_{\epsilon};L^2(\mathbf{K})\right)$ is solution of (\ref{ekcefifika}). Therefore
the Cauchy problem for (\ref{ekcefifika}) is locally well posed. To
obtain the global existence it is sufficient to prove that there
exists $C>0$ such that
\begin{equation}
 \label{estiKRAKa}
 \Vert\tilde{\psi}(\tau)\Vert_{H^1}^2+\Vert\partial_{\tau}\tilde{\psi}(\tau)\Vert_{L^2}^2\leq C\left[\Vert \tilde{\psi}(\tau_0)\Vert^2_{H^1}+\Vert \tilde{\psi}(\tau_0)\Vert^4_{H^1}+\Vert
  \partial_{\tau}\tilde{\psi}(\tau_0)\Vert_{L^2}^2\right]\exp\left(\left\vert\int_{\tau_0}^{\tau}(1+\mid\tilde{A}(\sigma)\mid)d\sigma\right\vert\right)
\end{equation}
This energy estimate is easily obtained by repeating the proof of
(\ref{estiKRAK}). Now we can define $\check{u}(\tau)$ for
$\tau\in[0,h]$ by the formula
$$
\check{u}(\tau)=\check{\Omega}^{-1}(\tau)e^{-\int_0^{\tau}\check{A}(s)ds}\tilde{\psi}(\tau).
$$
Finally we extend $\check{u}(\tau)$ for $\tau\in[h,\tau_+]$ by solving,
with Proposition \ref{propconl},
  the Cauchy problem for (\ref{nlkg}) with initial data given at
  $\tau_0=h$.

We now show that given $\tau_1,\tau_2\in[-h,h]$, the
bijection
$(\tilde{\psi}(\tau_1),\partial_{\tau}\tilde{\psi}(\tau_1))\mapsto
(\tilde{\psi}(\tau_2),\partial_{\tau}\tilde{\psi}(\tau_2))$ is
Lipschitz. We deduce from (\ref{zekinkin}) that given two solutions
$\tilde{\psi}$ and $\tilde{\psi}_*$, we have
\begin{equation*}
  \begin{split}
 \left\Vert
\left(
   \begin{array}{c}
     \tilde{\psi}\\
     \partial_{\tau}\tilde{\psi}
   \end{array}
 \right)(\tau_2)
 -
 \left(
   \begin{array}{c}
     \tilde{\psi}_*\\
     \partial_{\tau}\tilde{\psi}_*
   \end{array}
 \right)(\tau_2)
\right\Vert_{H^1\times L^2}
&\leq
 \left\Vert
\left(
   \begin{array}{c}
     \tilde{\psi}\\
     \partial_{\tau}\tilde{\psi}
   \end{array}
 \right)(\tau_1)
 -
 \left(
   \begin{array}{c}
     \tilde{\psi}_*\\
     \partial_{\tau}\tilde{\psi}_*
   \end{array}
 \right)(\tau_1)
\right\Vert_{H^1\times L^2}\\
&+
2\left\vert\int_{\tau_1}^{\tau_2}M\left(1+\Vert\tilde{\psi}(\tau)\Vert^2_{H^1}+\Vert\tilde{\psi}_*(\tau)\Vert^2_{H^1}\right)
  \Vert\tilde{\psi}(\tau)-\tilde{\psi}_*(\tau)\Vert_{H^1}d\tau\right\vert\\
&+
2\left\vert\int_{\tau_1}^{\tau_2}\mid\tilde{A}(\tau)\mid
  \Vert\partial_{\tau}\tilde{\psi}(\tau)-\partial_{\tau}\tilde{\psi}_*(\tau)\Vert_{L^2}d\tau\right\vert.
\end{split}
\end{equation*}
We apply the Gronwall lemma and (\ref{estiKRAKa}) to conclude there
exists a continuous function $\mathfrak{K}$ independent of
$\tilde{\psi}$ and $\tilde{\psi}_*$, such that for any
$\tau_j\in[-h,h]$ we have
\begin{equation*}
  \begin{split}
 \left\Vert
\left(
   \begin{array}{c}
     \tilde{\psi}\\
     \partial_{\tau}\tilde{\psi}
   \end{array}
 \right)(\tau_2)
 -
 \left(
   \begin{array}{c}
     \tilde{\psi}_*\\
     \partial_{\tau}\tilde{\psi}_*
   \end{array}
 \right)(\tau_2)
\right\Vert_{H^1\times L^2}
\leq
&
\mathfrak{K}\left(
  \left\Vert
\left(
   \begin{array}{c}
     \tilde{\psi}\\
     \partial_{\tau}\tilde{\psi}
   \end{array}
 \right)(\tau_1) \right\Vert_{H^1\times L^2}
,
\left\Vert
 \left(
   \begin{array}{c}
     \tilde{\psi}_*\\
     \partial_{\tau}\tilde{\psi}_*
   \end{array}
 \right)(\tau_1)
\right\Vert_{H^1\times L^2}
\right)\\
&\times
\left\Vert
\left(
   \begin{array}{c}
     \tilde{\psi}\\
     \partial_{\tau}\tilde{\psi}
   \end{array}
 \right)(\tau_1)
 -
 \left(
   \begin{array}{c}
     \tilde{\psi}_*\\
     \partial_{\tau}\tilde{\psi}_*
   \end{array}
 \right)(\tau_1)
\right\Vert_{H^1\times L^2}
\end{split}
\end{equation*}

Finally we can see that $\mathfrak{S}$ is a bi-Lipschitz bijection
  as a composition of bi-Lipschitz bijections:
  $
  \left(\hat{u}(\tau_-),\partial_{\tau}\hat{u}(\tau_-)\right)
  \mapsto
  \left(\hat{u}(-h),\partial_{\tau}\hat{u}(-h)\right)
  \mapsto
  \left(\tilde{\psi}(-h),\partial_{\tau}\tilde{\psi}(-h)\right)
  \mapsto
  \left(\tilde{\psi}(h),\partial_{\tau}\tilde{\psi}(h)\right)
  \mapsto
  \left(\check{u}(h),\partial_{\tau}\check{u}(h)\right)
  \mapsto
  \left(\check{u}(\tau_+),\partial_{\tau}\check{u}(\tau_+)\right).
  $

  \fin

  The non-linearity $\bar{f}=-\kappa\mid \bar{u}\mid^2\bar{u}$ is classic in
  Quantum Field Theory but in General Relativity, the non-linearities
  of the wave equations depend on derivatives of the unknown. If
  $\bar{f}=Q(x,\bar{u},\nabla\bar{u},\nabla^2\bar{u})$, the Liouville
  transform leads to (\ref{KG}) with a strongly singular non-linearity due to the
  derivatives of $\bar{\Omega}$ since
  $\bar{f}=Q_1(x,\bar{\Omega},\nabla\bar{\Omega},
  \nabla^2\bar{\Omega},\varphi,
  \nabla\bar{\varphi},\nabla^2\bar{\varphi})$. We known that in
  similar situations, the Fuchsian analysis is a powerful tool used in
  General Relativity (see {\it e.g.} \cite{beyer2010},
  \cite{beyer2019}, \cite{fournodavlos}, \cite{rendall}), hence we could expect that for some
  non-linearities $\bar{f}$ involving derivatives of the field, 
  (\ref{KG}) has a ``Fuchsian structure'' and these methods could be applied.




\begin{thebibliography}{10}

\bibitem{RIP}
A. Bachelot,
Wave asymptotics at a time cosmological singularity: classical and
quantum scalar field,
{\it Comm. Math. Phys.}, 369 (2019), 973--1020.

\bibitem{dodeca}
A. Bachelot, A. Bachelot-Motet,
Waves on accelerating dodecahedral universes,
{\it  Class. Quantum Grav. } 34   (2017),  no. 5, 055010, 39 pp.

\bibitem{battefeld}
D. Battefeld, P. Peter,
A critical review of classical bouncing cosmologies,
{\it Phys. Rep.} 571 (2015), 1--66. 

\bibitem{beyer2010}
F. Beyer, P. G. LeFloch,
Second-order hyperbolic Fuchsian systems and applications,
{\it  Class. Quantum Grav. } 27 (2010), 245012.

\bibitem{beyer2019}
 F. Beyer, T. A. Oliynyk, J. A. Olvera-Santamar{\`{\i}}a,
 The Fuchsian approach to global existence for hyperbolic equations, (2019),
  arXiv:1907.04071

\bibitem{brandenberger}
 R. Brandenberger, P. Peter,
 Bouncing Cosmologies: Progress and Problems,
 {\it Found. Phys.} 47 (2017) no 6, 797--850.

\bibitem{cagnac}
F. Cagnac, Y. Choquet-Bruhat,
Solution globale d'une \'equation non lin\'eaire sur une vari\'et\'e
hyperbolique,
{\it  J. Math. Pures Appl.} (9), 63 (1984), 377--390.

\bibitem{camara}
U. Camara da Silva, A. L. Alves Lima, G. M. Sotkov,
Scale factor duality for conformal cyclic cosmologies,
{\it JHEP} 11 (2016), 090.

\bibitem{delsanto}
D. Del Santo, T. Kinoshita, M. Reissig,
Klein-Gordon Type Equations with a Singular Time-dependent Potential,
{\it Rend. Istit. Mat. Univ. Trieste} 39 (2007), no. 11--12, 141--175.

\bibitem{ebert2017}
M. R. Ebert, W. N. Nascimento,
A classification for wave models with time-dependent mass and speed of
propagation,
{\it Adv. Differential Equations} 23 (2018), 847--888.

\bibitem{ebert2018}
M. R. Ebert, M. Reissig,
Regularity theory and global existence of small data solutions to
semi-linear de Sitter models with power non-linearity,
{\it Nonlinear Analysis: Real World Appl.} 40 (2018), 14--54.

\bibitem{fournodavlos}
G. Fournodavlos, J. Luk,
Asymptotically Kasner-like singularities, arXiv:2003.13591.

\bibitem{fried}
H. Friedrich,
Smooth Non-Zero Rest-Mass Evolution Across Time-Like Infinity,
{\it Ann. Henri Poincar\'e} 16 (2015), 2215--2238.

\bibitem{fried2017}
H. Friedrich,
Sharp Asymptotics for Einstein-$\lambda$-Dust Flows,
{\it Comm. Math. Phys.} 350 (2017), 803--844.

\bibitem{galstian2015}
A. Galstian, K. Yagdjian,
Global solutions for semilinear Klein-Gordon equations in FLRW
spacetimes,
{\it Nonlinear Analysis} 113 (2015), 339--356.

\bibitem{galstian2017}
A. Galstian, K. Yagdjian,
Global in time existence of self-interacting scalar  field in de Sitter
spacetimes,
{\it Nonlinear Analysis: Real World Appl.} 34 (2017), 110--139.

\bibitem{gasperini}
 M. Gasperini, G. Veneziano,
 The Pre--Big Bang Scenario in String Cosmology,
{\it Phys. Rep.} 373 (2003), 1--212.
  
\bibitem{joudioux}
J. Joudioux,
Conformal scattering for a nonlinear wave equation,
{\it J. Hyperbolic Differ. Equ.} 9 (2012), no. 1, 1--65.

\bibitem{klein}
D. Klein, J. Reschke,
Pre-big Bang Geometric Extensions of Inflationary Cosmologies,
{\it Ann. Henri Poincaré} 19 (2018), no. 2, 565--606.

\bibitem{lubbe}
C. L\"ubbe,
Conformal scalar fields, isotropic singularities and conformal cyclic
cosmologies, (2013),
arXiv:1312.2059 [gr-qc]

\bibitem{meissner}
 K. A. Meissner, P. Nurowski,
 Conformal transformations and the beginning of the Universe,
 {\it Phys. Rev.D } 95 (2017), no. 8, 084016 .

\bibitem{nakamura}
M. Nakamura,
The Cauchy problem for semi-linear Klein-Gordon equations in de Sitter spacetime,
{\it J. Math. Anal. Appl.} 410 (2014), 445--454.

\bibitem{newman}
 E. Newman,
 A fundamental solution to the CCC equations,
 {\it  Gen. Rel. Grav.} 46 (2014), 1717.

\bibitem{nicolas1995}
 J.-P. Nicolas,
 Non linear Klein-Gordon equation on Schwarzschild-like metrics,
 {\it J. Math. Pures Appl.} 74 (1995), 35--58.

\bibitem{nicolas2002}
 J.-P. Nicolas,
 A nonlinear Klein-Gordon equation on Kerr metrics,
 {\it J. Math. Pures Appl.} 81 (2002), no. 9, 885--914.

\bibitem{ccc}
R. Penrose,
{\it Cycles of Time: An Extraordinary New View of the Universe}, 
(Bodley
Head, 2010).

\bibitem{penrose2018}
 R. Penrose,
 The Big Bang and its Dark-Matter Content: Whence, Whither, and
 Wherefore,
 {\it Found. Phys.} 48 (2018), 1177--1190.

\bibitem{rendall}
A. D. Rendall,
Fuchsian analysis of singularities in Gowdy
spacetimes beyond analyticity,
{\it Classical Quantum Gravity} 17 (2000),
no. 16, 3305--3316.

\bibitem{ringstrom2008}
H. Ringstr\"om,
Future stability of the Einstein non-linear scalar field system,
{\it Invent. Math.} 173 (2008), 123--208.

\bibitem{ringstrom2013}
H. Ringstr\"om,
{\it On the topology and future stability of the universe},
 Oxford Mathematical Monographs (Oxford University Press,  2013).

\bibitem{ringstrom2017}
H. Ringstr\"om,
Linear systems of wave equations on cosmological backgrounds with
convergent asymptotics,
{\it preprint} (2017), arXiv:1707.02803.

\bibitem{ringstrom2019}
H. Ringstr\"om,
A Unified Approach to the Klein-Gordon Equation on Bianchi
Backgrounds,
{\it Comm. Math. Phys.} 372 (2019),no.  2, 599--656.

\bibitem{rodnianski2018}
I. Rodnianski,  J. Speck,
A regime of linear stability for the Einstein-scalar field system with
applications to nonlinear big bang formation,
{\it Ann. of Math.} (2) 187 (2018), no. 1, 65--156.

\bibitem{rodnianski2018-1}
I. Rodnianski,  J. Speck,
Stable big bang formation in near-FLRW solutions to the Einstein-scalar field and Einstein-stiff fluid systems,
{\it Selecta Math. (N.S.)} 24 (2018), no. 5, 4293--4459.

\bibitem{strauss}
W. A. Strauss,
On continuity of functions with values in various Banach spaces,
{\it Pacific J. Math.} 19 (3) (1966), 543--551.

\bibitem{teschl}
 G. Teschl,
 {\it Ordinary Differential Equations and Dynamical Systems},
 Graduate Studies in Mathematics, Volume 140 (Amer. Math. Soc., Providence, 2012).

\bibitem{tod}
P. Tod,
The equations of conformal cyclic cosmology,
{\it Gen. Relativ. Gravit.} 47, (2015) no. 3, Art. 17, 13 pp.

\bibitem{vasy}
A. Vasy,
The wave equation on asymptotically de Sitter-like spaces,
{\it Adv. Math.} 223 (2010), 49--97.


\end{thebibliography}
\end{document}